\documentclass[a4paper]{article}

\usepackage{INTERSPEECH2018}
\usepackage{stackengine}
\usepackage{multirow}
\newcommand\IncG[2][]{\addstackgap{%
		\raisebox{-.5\height}{\includegraphics[#1]{#2}}}}

\title{DNN Based Speech Enhancement for Unseen Noises Using Monte Carlo Dropout}
\name{Nazreen P.M., A. G. Ramakrishnan}
\address{
MILE Lab, Department of Electrical Engineering, Indian Institute of Science, Bangalore 560012}
\email{nazreenp@iisc.ac.in, agr@iisc.ac.in}

\begin{document}

\maketitle
\begin{abstract}
In this work, we propose the use of dropouts as a Bayesian estimator for increasing the generalizability of a deep neural network (DNN) for speech enhancement. By using Monte Carlo (MC) dropout, we show that the DNN performs better enhancement in unseen noise and SNR conditions. The DNN is trained on speech corrupted with Factory2, M109, Babble, Leopard and Volvo noises at SNRs of 0, 5 and 10 dB and tested on speech with white, pink and factory1 noises. Speech samples are obtained from the TIMIT database and noises from NOISEX-92. In another experiment, we train five DNN models separately on speech corrupted with Factory2, M109, Babble, Leopard and Volvo noises, at 0, 5 and 10 dB SNRs. The model precision (estimated using MC dropout) is used as a proxy for squared error to dynamically select the best of the DNN models based on their performance on each frame of test data.
\footnote{Submitted on 23 March 2018 for Interspeech 2018}       
\end{abstract}

\section{Introduction}
 \vspace{-0.2em}
Single channel speech enhancement has been a challenging problem for decades. Speech enhancement techniques find several applications such as automatic speech recognition, hearing aids and speaker recognition.
Methods proposed in the past include unsupervised methods such as spectral subtraction \cite{ss,GA}, Wiener filtering \cite{wiener}, minimum mean-square error estimators \cite{ephraim1984speech}, estimators based on Gaussian prior distributions \cite{martin2005speech, erkelens2007minimum} and residual-weighting schemes \cite{yegnanarayana1999speech, jin2006speech, prasanna2011}. Most of these methods may perform poorly when the background noise is non-stationary and in unexpected acoustic conditions.

In supervised learning methods, prior information is fed into the models and hence they are expected to  perform better than unsupervised methods \cite{srinivasan2006codebook, ephraim1992bayesian, sameti1998hmm}. Neural networks have been shown to learn the mapping between noisy and clean speech \cite{tamura1989analysis, xie1994family, wan1999networks}. However, these models are small networks with a single hidden layer and cannot fully learn the mapping. Deep architectures have conquered this area recently, since these networks with multiple layers have been shown to better learn the complex mapping between noisy and clean features and hence give really good enhancement performances. Hinton \textit{et al.} proposed a greedy layer-wise unsupervised learning algorithm \cite{hinton2006reducing, hinton2006fast}. Mass et al. \cite{maas2012recurrent} use deep recurrent neural networks for feature enhancement for noise robust ASRs. 

One of the major issues encountered by deep neural network (DNN) based enhancement is the degradation of performance for  noises unseen during training. The model learns the mapping between noisy and clean speech well for those noises and signal to noise ratios (SNRs) with which it is trained, but performs poorly on speech corrupted by an unseen noise. In fact, this itself could be dealt with as a challenging task in speech enhancement scenario. Though not dealt with separately, techniques have been proposed in the past to address this problem. In \cite{xu2014experimental}, they have proposed a regression DNN-based speech enhancement framework, where they train a wide neural network using a really huge collection of data of about 100 hours of various noise types. In \cite{wang2013towards}, a DNN-SVM based system is trained on a variety of acoustic data for a huge amount of time. A noise aware training technique is adopted in \cite{xu2015regression}, where a noise estimate is appended to the input feature for training. They use about 2500 hours of data for training the network. 

Hinton~\cite{dropout, srivastava2014dropout} introduced the concept of dropouts to reduce overfitting during DNN training. Though dropout omits weights during training, it is inactive during the inference stage, whereby all the neurons contribute to the prediction.

Gal and Ghahramani \cite{gal2016dropout} proposed using dropouts during testing, by showing a theoretical relationship between dropout and approximate inference in a Gaussian process. In \cite{kendall2016modelling}, they show that by enabling dropouts during testing, and averaging the results of multiple stochastic forward passes, the predictions usually become better. They refer to this technique as Monte Carlo (MC) dropouts, where the output samples are MC samples from the posterior distribution of models. In \cite{gal2016dropout}, they show that the model uncertainty can also be estimated from these samples.

In this work, we explore how to use the idea of MC dropout  to improve the generalizability of speech models, thereby improving the enhancement performance in a mismatched condition. We show that when the input is a noisy speech corrupted with an unseen noise, the use of MC dropout instead of normal dropout can give a better output. Hence the same concept could be applied to any of the above mentioned DNN speech models to further improve the generalizability of the output to get a better performance during unseen noise scenarios.


We also explore the usage of model uncertainty in problems where multiple noise specific DNN models are used. By using model uncertainty as an estimate of the prediction error for a sample, this technique can enable the selection of the model with the least prediction error on a frame by frame basis. A similar approach of selecting the best model based on an error estimate is proposed in \cite{papadopoulos2016long} for robust SNR estimation. They trained a separate DNN as a classifier to select a particular regression model for SNR estimation. However, this approach does not ameliorate the original problem of mismatch in training and testing conditions. In our proposed algorithm, we use the intrinsic uncertainty of a model to estimate the prediction error. Since this method extracts information from the model itself, it has the potential to be a better representative of the prediction error. Our method also circumvents the issue of unseen testing conditions, since according to \cite{gal2016dropout}, the model uncertainty itself is an indicator of unseen data.

Our approach to improve the generalization involves two methods. In the first approach, we show that MC dropout estimate shows improvement in the generalization performance of DNN and apply this to speech enhancement. We train two DNN models, one without MC dropout and the other using MC dropout, with speech corrupted with five different noises at SNRs 0, 5 and 10 dB. During testing, multiple repetitions are performed by dropping out different units every time and the empirical mean of all the outputs are taken. We have analysed that using MC dropout shows promise in improving the enhancement performance for unseen noise and SNR scenarios.

The second approach is an analysis on the use of model uncertainty as an estimate of the prediction error for a sample, where multiple DNN models are used. Five DNN models are trained, each using MC dropout. Each model is trained on noisy speech corrupted with a different noise at SNRs of 0, 5 and 10 dB. In a general scenario, one needs to identify the noise type to choose the right noise model to enhance the input noisy speech. However, in the case of an unseen noise scenario, the selection of the appropriate model becomes tricky. In such cases, we need to ensure that the chosen model is the one that gives the lowest error and hence a better enhancement performance. We use the model uncertainty, estimated from the output samples of each model, as an estimate of the prediction error and choose the model based on it. Our experiments show that the stronger the correlation between the model uncertainty and the squared error, the better is the enhancement performance.

\vspace{-0.1cm}

\section{DNN based speech enhancement}
\vspace{-0.1cm}
Under additive model, the noisy speech can be represented as,
\vspace{-0.1 cm}
\begin{equation}
	x_t(m)=s_t(m)+n_t(m)
\end{equation}
\vspace{-0.1 cm}
where $ x_t(m) $, $ s_t(m) $ and $ n_t(m) $ are the $m^{th}$ samples of the noisy speech, clean speech and noise signal, respectively. Taking the short time Fourier transform (STFT), we have,
\vspace{-0.1 cm}
\begin{equation}
	x(\omega_k)=s(\omega_k)+n(\omega_k)
\end{equation}
where $\omega_k = (2 \pi k/R)$, $k=0,1,2...R-1$, $k$ is the index and $R$ is the number of frequency bins. Taking the magnitude of the STFT, the noisy speech can be approximated as
\begin{equation}
X \approx S+N \in \mathbb{R}^{R\times 1} 
\end{equation}
\vspace{-0.1cm}
where $ S $ and $N$ represent the spectra of the clean speech and the noise, respectively. 

A DNN based regression model is trained using the magnitude STFT features of clean and noisy speech. The noisy features are then fed to this trained DNN to predict the enhanced features, $\hat{S}$. The enhanced speech signal is obtained by using the inverse Fourier transform of $\hat{S}$ with the phase of the noisy speech signal and overlap-add method. 
\vspace{-0.1cm}
\subsection{Basic DNN architecture}
\vspace{-0.1cm}
The proposed baseline system uses a DNN to learn the complex mapping of input noisy speech to clean speech. It consists of 3 fully connected layers of 2048 neurons and an output layer of 257. We use ReLU non-linearity as the activation function in all the 3 layers. Our output activation is also ReLU to account for the nonnegative nature of STFT magnitude. Backpropagation algorithm is used for training. Stochastic gradient descent is used to minimize the mean square logarithmic error ($E_{r}$) between the noisy and clean magnitude spectra:
\vspace{-0.1cm}
\begin{equation}
 E_{r} =\frac{1}{R} \sum_{k=1}^{R}(log(S(k)+1)-log(\hat{S(k)}+1))^2
 \vspace{-0.1cm}
\end{equation}
\vspace{-0.1cm}
where $\hat{S}$ and $S$ denote the estimated and reference spectral features, respectively, at sample index $k$. This baseline system is only meant to illustrate the usage of our system. Consequently, we do not use the incremental improvements in the literature.
\vspace{-0.2cm}
\section{Proposed methods for generalized speech models}
\vspace{-0.2cm}
Gal and Ghahramani \cite{gal2016dropout} have shown a theoretical relationship between dropout \cite{dropout} and approximate inference in a Gaussian process. The proposed system augments the baseline system by dropouts as a bayesian approximation. By using this approximation, a distribution over the weights is learnt, thereby giving uncertainty of the output.

The network output is simulated with input $X$, using dropout same as that employed during the training time. During testing, $T$ repetitions are performed, with different random units in the network dropped out every time, obtaining the results $ \{\hat{S_{t}(X)}\}; 1\leq t\leq T$ . It is shown in \cite{gal2016dropout} that averaging forward passes through the network is equivalent to Monte Carlo integration over a Gaussian process posterior approximation. Empirical estimators of the predictive mean ($E(S)$) and variance (uncertainty, $V(S)$) from these samples are given as:

\vspace{-0.2cm}
\begin{equation}
E(S) \approx \frac{1}{T}\sum_{t=1}^{T} \hat{S_{t}(X)}
\label{eq5}
\end{equation}
\vspace{-0.2cm}
\begin{equation}
V(S)\approx \tau ^{-1} I_D + \frac{1}{T}\sum_{t=1}^{T} \hat{S_{t}(X)}^T \hat{S_{t}(X)} - E(S)^T E(S)
\label{eq6}
\end{equation}
where $\tau = {l^2 p}/{2N \lambda}$ ; $l$: defined prior length scale, $p$: probability of the units not being dropped, $N$: total input samples,  $\lambda$: regularisation weight decay, which is zero for our experiments.

\vspace{-0.2cm}
\subsection{Single DNN model using MC dropout}
\vspace{-0.4cm}
\begin{figure}[ht!]%
	
	
	\centering

	\includegraphics[height=1.8cm, width=7.5cm]{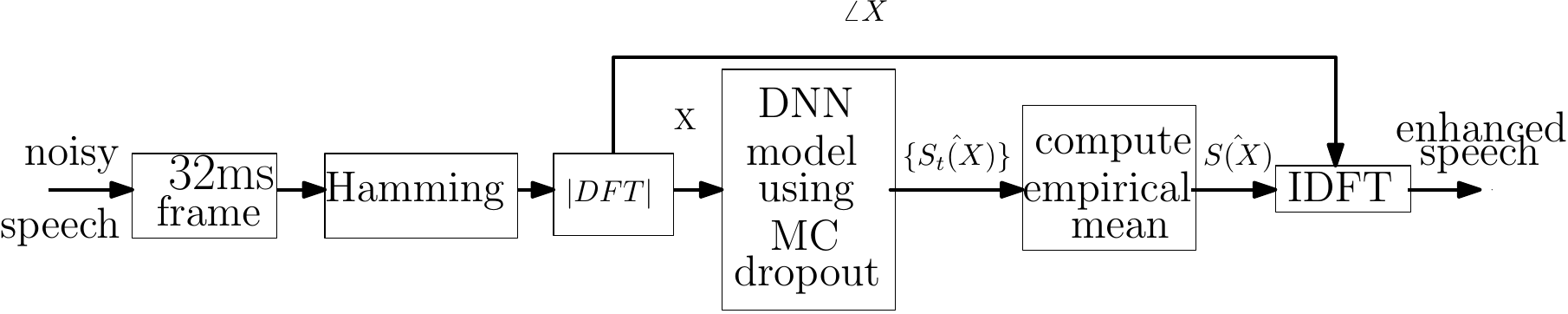}


	\caption{Enhancement using single DNN-MC dropout model.}
		\label{fig1}	
	\vspace{-0.2 cm}
\end{figure}
A single DNN model is trained using MC dropout with speech corrupted with five noises: factory2, m109, leopard, babble and volvo at SNRs 0, 5 and 10 dB. A baseline model is trained on the same noises and SNRs, without the MC dropout. 
Figure \ref{fig1} shows the block diagram of the proposed approach. During testing of MC dropout model, given a noisy speech frame $X$, multiple repetitions are performed by dropping out different units each time giving $T$ different outputs, $ \{\hat{S_{t}(X)}\}; 1\leq t\leq T$. The empirical mean of these outputs are used as the estimated output $\hat{S(X)}$  as shown in eqn. \ref{eq5}. Enhanced speech is obtained as the inverse Fourier transform of $\hat{S(X)}$ with the phase of the noisy speech signal and overlap-add method.

\subsection{Multiple DNN models using MC dropout with predictive variance (model uncertainty) as the selection scheme}
\vspace{-0.2cm}
\begin{figure}[ht!]%
	
	
	\centering

	\includegraphics[height=3.cm, width=7.5cm]{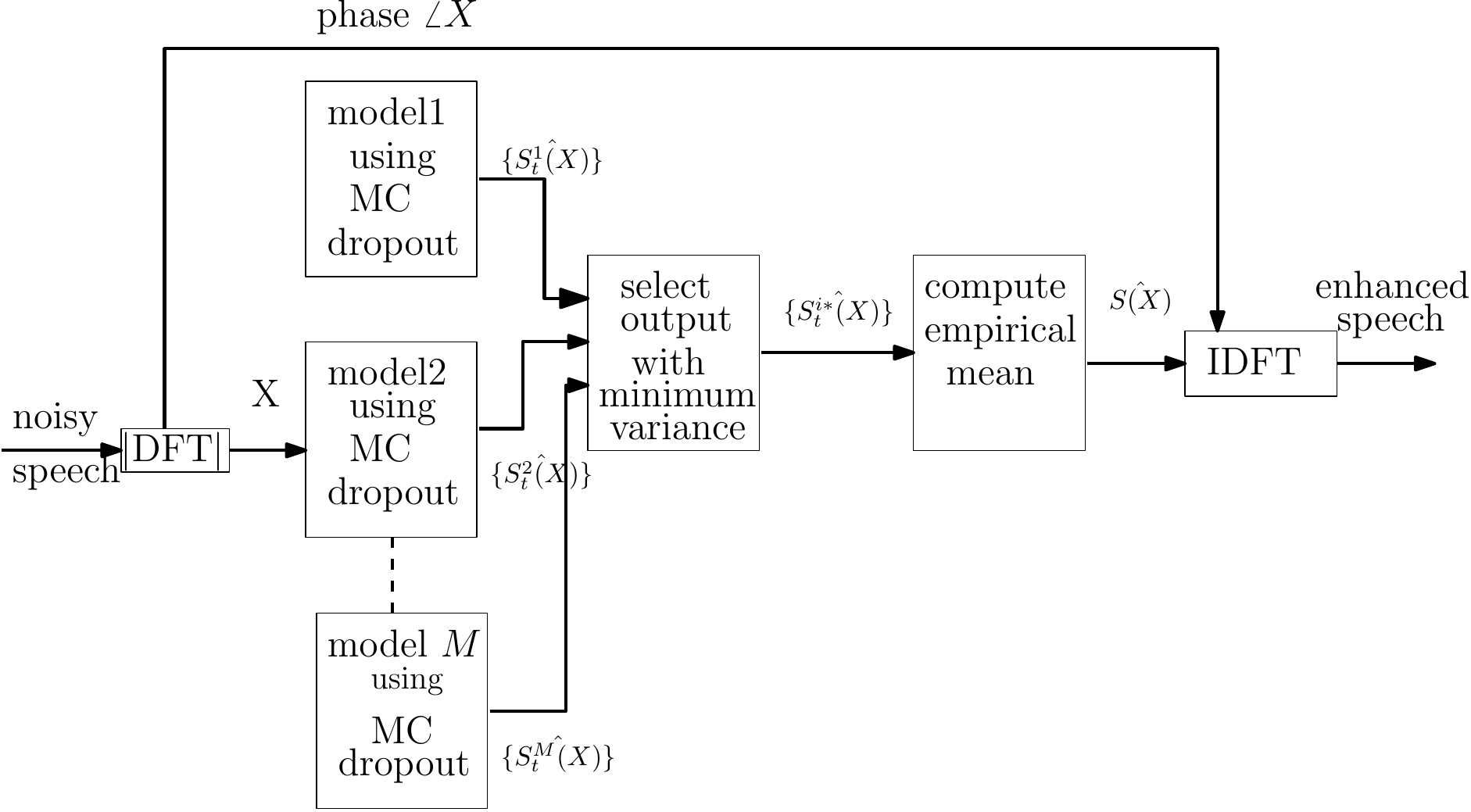}


	\caption{Enhancement using multiple DNN-MC dropout models with predictive variance as the selection criteria.}
		\label{fig2}	
	\vspace{-0.4 cm}
\end{figure}

Model-specific enhancement techniques depend on a model selector \cite{dnn_phn, my}, which ensures that the model chosen for enhancing each frame  entails an overall improved performance. Given multiple noise-specific DNN models for enhancing a frame of noisy speech, one method to select the appropriate model is to first detect the type of noise. However, if speech has been corrupted with an unseen noise, the selection of the appropriate model gets harder, since the noise detector assumes that one of the models is trained with the correct noise.

In this work, we follow \cite{gal2016dropout} and say that since model uncertainty gives the intrinsic uncertainty of the model for a particular input, we can use it as an estimate of model error.   Given that this relation holds, we can build a framework as per Figure \ref{fig2}  to enhance speech. Thus, this approach works only when there is a strong correlation between model uncertainty and output error.

Figure \ref{fig2} shows the block diagram of the proposed approach. Five DNN-MC dropout models are trained on speech corrupted with factory2, leopard, m109, babble and volvo noises of SNRs 0, 5 and 10 dB. The architecture is the same as the one defined in Sec. 2.1. For a given noisy input frame $X$, each of these models generates an output by dropping out random units. $T$ repetitions are performed by each model by dropping different units every time, obtaining results $ \{\hat{S^i_{t}(X)}\}; 1\leq t\leq T ; 1\leq i\leq M$ ; where $i$ is the model index and $M=5$. The predictive variance (uncertainty) is computed for each of the $M$ different results. The model with the minimum variance is selected as the best one for that frame. The enhanced output $\hat{S}$ is estimated as the empirical mean of the $T$ outputs: $ \{\hat{S^{i*}_{t}(X)}\}; 1\leq t\leq T$. The enhanced speech signal is obtained as the inverse Fourier transform of $\hat{S}$ with the phase of the noisy speech signal and overlap-add method.

\vspace{-0.2cm}
\section{Experiments and Results}
\subsection{Experimental setup}
TIMIT \cite{timit} speech corpus is used for the study. The five noises used from the NOISEX-92 \cite{noisex} database are downsampled to 16 kHz to match the sampling rate of TIMIT, in order to synthesize noisy training and test speech signals. The magnitude STFT is computed on frames of size 32 ms with 10 ms frame shift, after applying Hamming window. A 512-point FFT is taken and we use only the first 257 points as input to the DNN, because of symmetry in the spectrum. A DNN based regression model is trained using the magnitude STFT features of clean and noisy speech. For multi-model MC dropout experiments, each DNN model is trained on one of the following noises: Factory 2, m109, leopard, babble and volvo each at SNRs 0, 5 and 10 dB. For the single model case, the DNN is trained on factory2, m109, leopard, babble and volvo noises at SNRs 0,5 and 10 dB for both baseline and MC dropout. During testing, the noisy features are fed to this trained DNN to predict the enhanced features, $\hat{S}$. The enhanced speech signal is obtained as the inverse Fourier transform of $\hat{S}$, using the phase of the noisy speech signal and overlap-add method. 

The DNN architecture used has been defined in Sec. 2.1. For our experiments, the number of repetitions $T$ is chosen as 50. The Adam optimizer \cite{kingma2014adam} is chosen, whose default regularization weight decay, $\lambda$ is zero and thus, $\tau^{-1} = 0$ in eqn.\ref{eq6}.
\vspace{-0.5cm}
\subsection{Results and discussion}
\begin{table*}[]
	\centering
	\caption{Performance evaluation of single DNN model with MC dropout}
	
	\vspace{-1em}
	
	\resizebox{0.99\linewidth}{!}{
		\begin{tabular}{|l|l|l|l|l|l|l|l|l|l|l|l|l|l|}
			\hline
			\textbf{}                     & \textbf{}     & \multicolumn{3}{l|}{\textbf{White}}                                                                          & \multicolumn{3}{l|}{\textbf{Pink}}                                                                           & \multicolumn{3}{l|}{\textbf{Factory1}}                                                                      & \multicolumn{3}{l|}{\textbf{Factory2}}                                 \\ \hline
			\textbf{SNR}                  & \textbf{Metric}     & \textbf{Noisy}                     & \textbf{Baseline}                  & \textbf{MC}                        & \textbf{Noisy}                     & \textbf{Baseline}                  & \textbf{MC}                        & \textbf{Noisy}                     & \textbf{Baseline}                  & \textbf{MC}                       & \textbf{Noisy}                     & \textbf{Baseline} & \textbf{MC}   \\ \hline
			\multirow{2}{*}{\textbf{-10}} & \textbf{SSE x10\textasciicircum 4}  & {3.64} & {3.36} & \textbf{3.14} & {3.96} & {0.874} & \textbf{0.848} & {3.69} & {0.720} & \textbf{0.70} & {4.13} & {0.0467}      & \textbf{0.0461}  \\ \cline{2-14} 
			& \textbf{SSNR} & {-8.9}                      & {-8.5}                      & \textbf{-8.4}                      & {-8.8}                      & {-6.7}                      & \textbf{-6.6}                      & {-8.7}                      & {-6.0}                      & \textbf{-5.9}                     & {-8.5}                      & {1.0}      & {1.0}  \\ \hline
			
			\multirow{2}{*}{\textbf{-5}} & \textbf{SSE x10\textasciicircum 4}  & {1.12} & {0.960} & \textbf{0.913} & {1.22} & {0.270} & \textbf{0.251} & {1.12} & {0.213} & \textbf{0.200} & {1.29} & {0.0198}      & \textbf{0.0197}  \\ \cline{2-14} 
			& \textbf{SSNR} & {-7.2}                      & {-6.6}                      & \textbf{-6.5}                      & {-7.1}                      & {-4.3}                      & \textbf{-4.2}                      & {-6.9}                      & {-3.51}                     & \textbf{-3.50}                     & {-6.7}                      & {3.05}     & {3.08} \\ \hline
			\multirow{2}{*}{\textbf{0}} & \textbf{SSE x10\textasciicircum 3}  & {3.41} & {2.81 } & \textbf{2.60} & {3.71} & {0.858}      & \textbf{0.843}  & {3.41} & {0.682}      & \textbf{0.671}   & {4.01} &{0.104}      & {0.104}  \\ \cline{2-14} 
			& \textbf{SSNR} & {-4.6}                      & {-3.9}                     & \textbf{-3.8}                      & {-4.5}                      & {-1.5}     & \textbf{-1.4} & {-4.4}                      & {-0.73}    & {-0.73} & {-4.1}                      & {5.1}      &{5.1}  \\ \hline
			
			\multirow{2}{*}{\textbf{5}} & \textbf{SSE x10\textasciicircum 3}  & 1.03 & 0.844               & \textbf{0.827}         & 1.12 & 0.291               & \textbf{0.288}         & 1.02 & 0.244               & \textbf{0.242}         & 1.24 & 0.069                & 0.069          \\ \cline{2-14} 
			& \textbf{SSNR} & -1.6                      & -0.7              & -0.7        & -1.4                      & 1.7               & 1.7         & -1.3                      & 2.2               & 2.2         & -0.9                      & 7.1               & 7.1         \\ \hline
			\multirow{2}{*}{\textbf{10}} & \textbf{SSE x10\textasciicircum 2}  & 3.08            & 2.70               & \textbf{2.67}         & 3.41            & 1.18               & \textbf{1.16}         & 3.09            & 1.07               & \textbf{1.06}         & 3.82            & 0.56                & 0.55          \\ \cline{2-14} 
			& \textbf{SSNR} & 2.0            & 2.7               & 2.7         & 2.2            & 4.7               & 4.7         & 2.3            & 5.0               & 5.0         & 2.6            & 8.9               & 8.9         \\ \hline
			
		\end{tabular}
	}
	
	\label{table1}
\end{table*}

Table \ref{table1} lists the improvements obtained in terms of sum squared error (SSE),  and segmental SNR (SSNR) \cite{hu2008evaluation} for single DNN-MC dropout model over the baseline for unseen noises. We use white, pink and factory 1 noises as unseen noises and factory2 as a seen noise. The reported results are the average over 100 files randomly selected from TIMIT \cite{timit}. The model achieves superior performance in most of the cases. It is to be noted that the improvement is significant for unseen noises like white noise, especially at low SNRs of -10 and -5 dB. Interestingly, the performance degrades with higher SNRs, though the model continues to perform better than the baseleine in terms of SSE. Though the proposed method does not result in significant improvement on seen noises, the performance is comparable to the baseline model. Hence, the observations validate the proposed method of using MC dropout to improve generalization performance on unseen noises.

\begin{table*}[]
	\centering
	\caption{Performance of multiple DNN-MC dropout models with predictive variance based selection on unseen noises.}
	\vspace{-1em}

	\resizebox{0.99\linewidth}{!}{
		\begin{tabular}{|l|l|l|l|l|l|l|l|l|l|l|l}
			\hline
			\textbf{}                     & \textbf{}     & \multicolumn{3}{l|}{\textbf{White}}                                                                          & \multicolumn{3}{l|}{\textbf{Pink}}                                                                           & \multicolumn{3}{l|}{\textbf{Factory1}}                                                                       \\ \hline
			\textbf{SNR}                  & \textbf{Metric}     & \textbf{Noisy}                     & \textbf{Baseline}                  & \textbf{MC}                        & \textbf{Noisy}                     & \textbf{Baseline}                  & \textbf{MC}                        & \textbf{Noisy}                     & \textbf{Baseline}                  & \textbf{MC}                        \\ \hline
			\multirow{2}{*}{\textbf{-10}} & \textbf{SSE x10\textasciicircum 4}  & {3.64} & {3.36} & \textbf{3.2} & {3.96} & {0.874} & \textbf{0.708} & {3.69} & {0.720} & \textbf{0.677}   \\ \cline{2-11} 
			& \textbf{SSNR} & {-8.9}                      & {-8.5}                      & \textbf{-8.4}                      & {-8.8}                      & {-6.7}                      & \textbf{-5.4}                      & {-8.7}                      & {-6.0}                      & \textbf{-5.3}                      \\ \hline
			
			\multirow{2}{*}{\textbf{-5}} & \textbf{SSE x10\textasciicircum 4}  & {1.12} & {0.960} & \textbf{0.936} & {1.22} & {0.270} & \textbf{0.261} & {1.12} & {0.213} & \textbf{0.20}   \\ \cline{2-11} 
			& \textbf{SSNR} & {-7.2}                      & {-6.6}                      & \textbf{-6.5}                      & {-7.1}                      & {-4.3}                      & \textbf{-3.7}                      & {-6.9}                      & {-3.51}                     & \textbf{-3.3}                       \\ \hline
			\multirow{2}{*}{\textbf{0}} & \textbf{SSE x10\textasciicircum 3}  & {3.41} & {2.81} & \textbf{2.70} & {3.71} & {0.858}      & {0.943}  & {3.41} & {0.682}      & {0.771}   \\ \cline{2-11} 
			& \textbf{SSNR} & {-4.6}                      & {-3.9}                     & \textbf{-3.8}                      & {-4.5}                      & {-1.5}     & \textbf{-1.3} & {-4.4}                      & {-0.73}    & {-0.83}   \\ \hline
			
			\multirow{2}{*}{\textbf{5}} & \textbf{SSE x10\textasciicircum 3}  & 1.03 & 0.844               & 0.857         & 1.12 & 0.291               & 0.391         & 1.02 & .244               & .285          \\ \cline{2-11} 
			& \textbf{SSNR} & -1.6                      & -0.7              & -0.7        & -1.4                      & 1.7               & 1.6         & -1.3                      & 2.2               & 2.0         \\ \hline
			\multirow{2}{*}{\textbf{10}} & \textbf{SSE x10\textasciicircum 2}  & 3.08            & 2.70               & 2.73         & 3.41            & 1.18               & 1.40         & 3.09            & 1.07               & 1.24         \\ \cline{2-11} 
			& \textbf{SSNR} & 2.0            & 2.7               & 2.7         & 2.2            & 4.7               & 4.5         & 2.3            & 5.0               & 4.8          \\ \hline
			
		\end{tabular}
		
	}
		\label{table2}
\end{table*}
\begin{figure*}
	
	\begin{tabular}{p{.05\textwidth}p{.18\textwidth}p{.18\textwidth}p{.18\textwidth}p{.18\textwidth}p{.18\textwidth}}
		& \hspace{1em}Factory2 model & \hspace{-.7em}Leopard model & \hspace{-1.5em}
		M109 model & \hspace{-2em}Babble model & \hspace{-3em}Volvo model\\
		-10dB  
		&\hspace{-1.5em}\IncG[width=.2\textwidth,height=.1\textwidth]{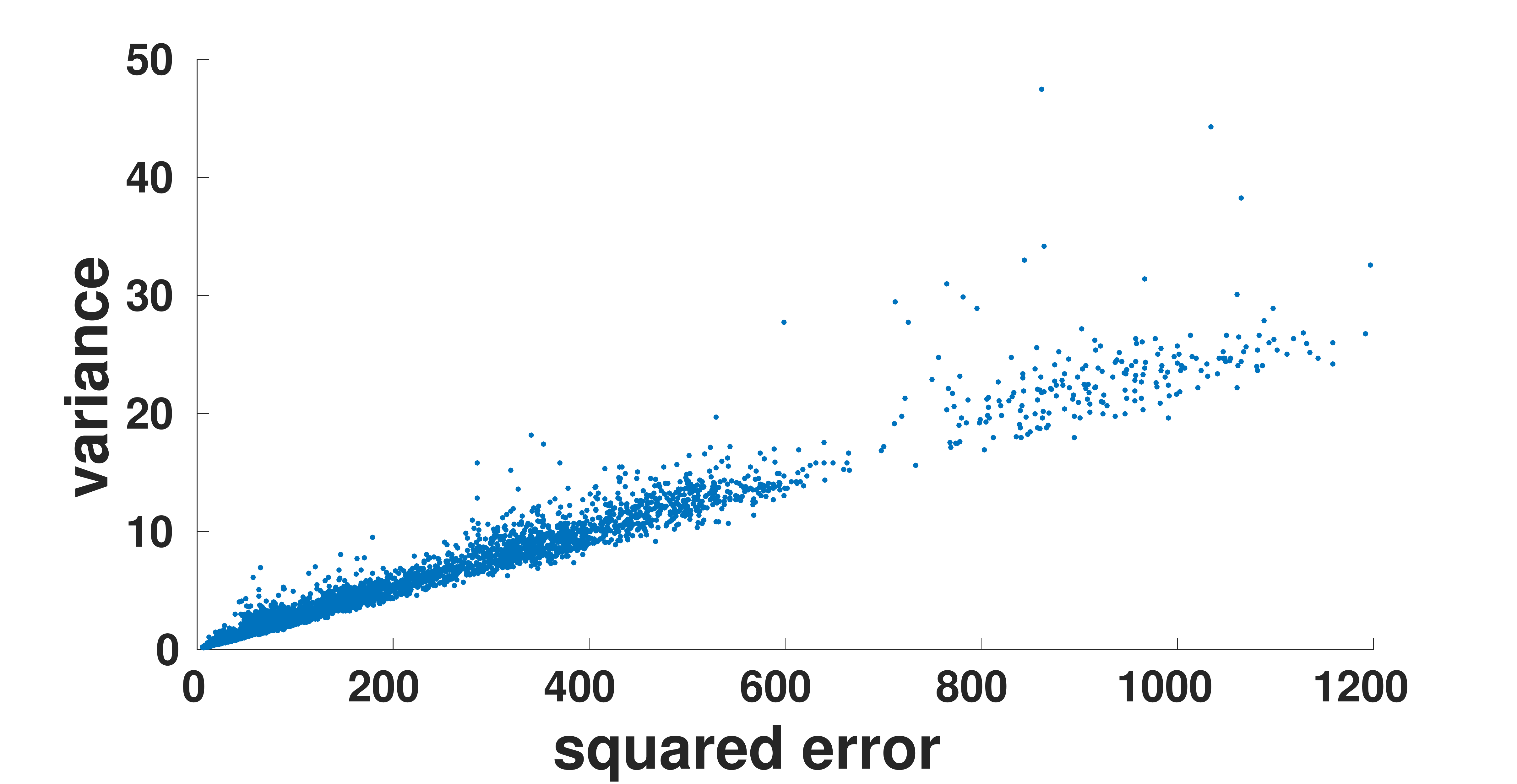}
		&\hspace{-3em}\IncG[width=.2\textwidth,height=.1\textwidth]{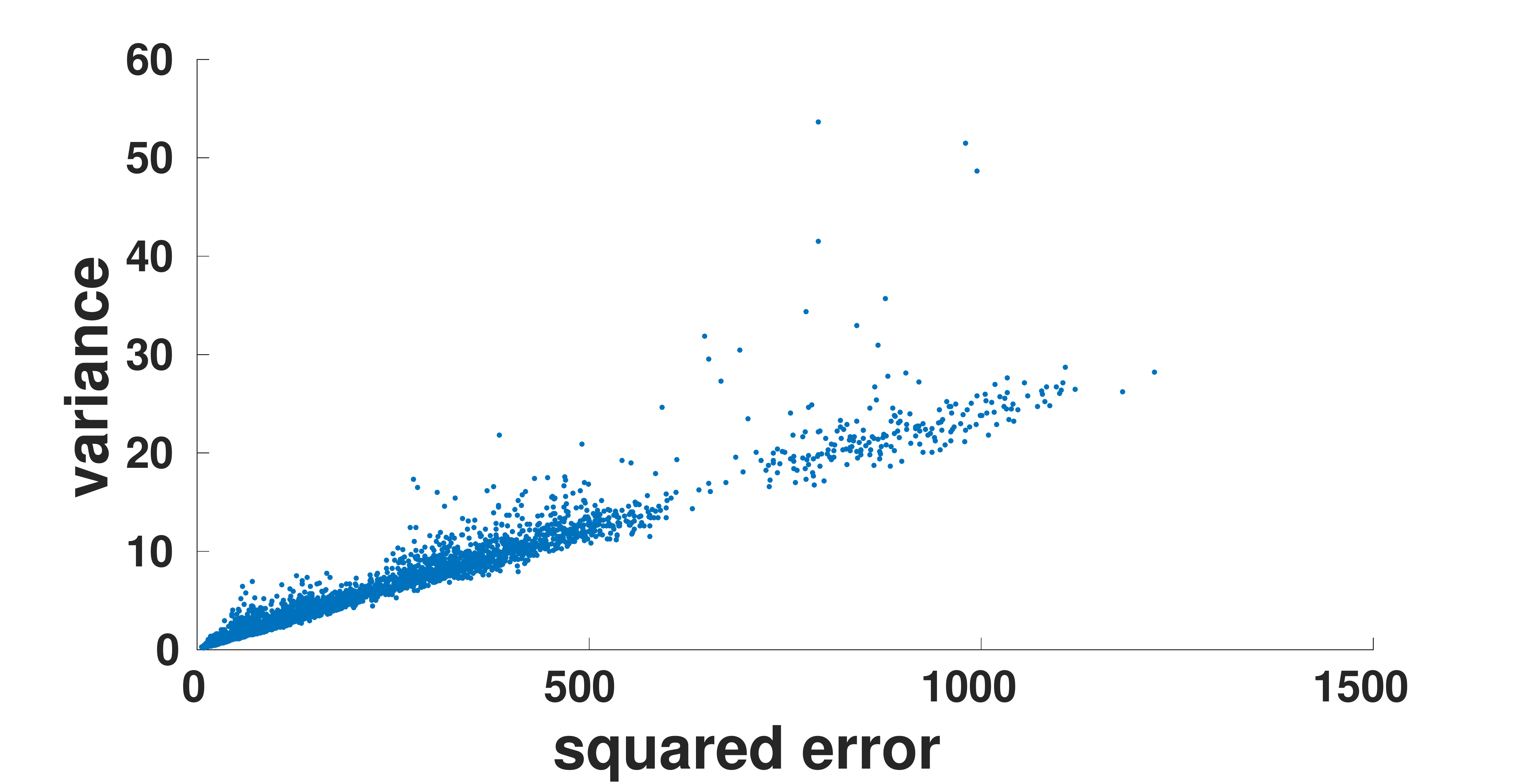}
		&\hspace{-4em}\IncG[width=.2\textwidth,height=.1\textwidth]{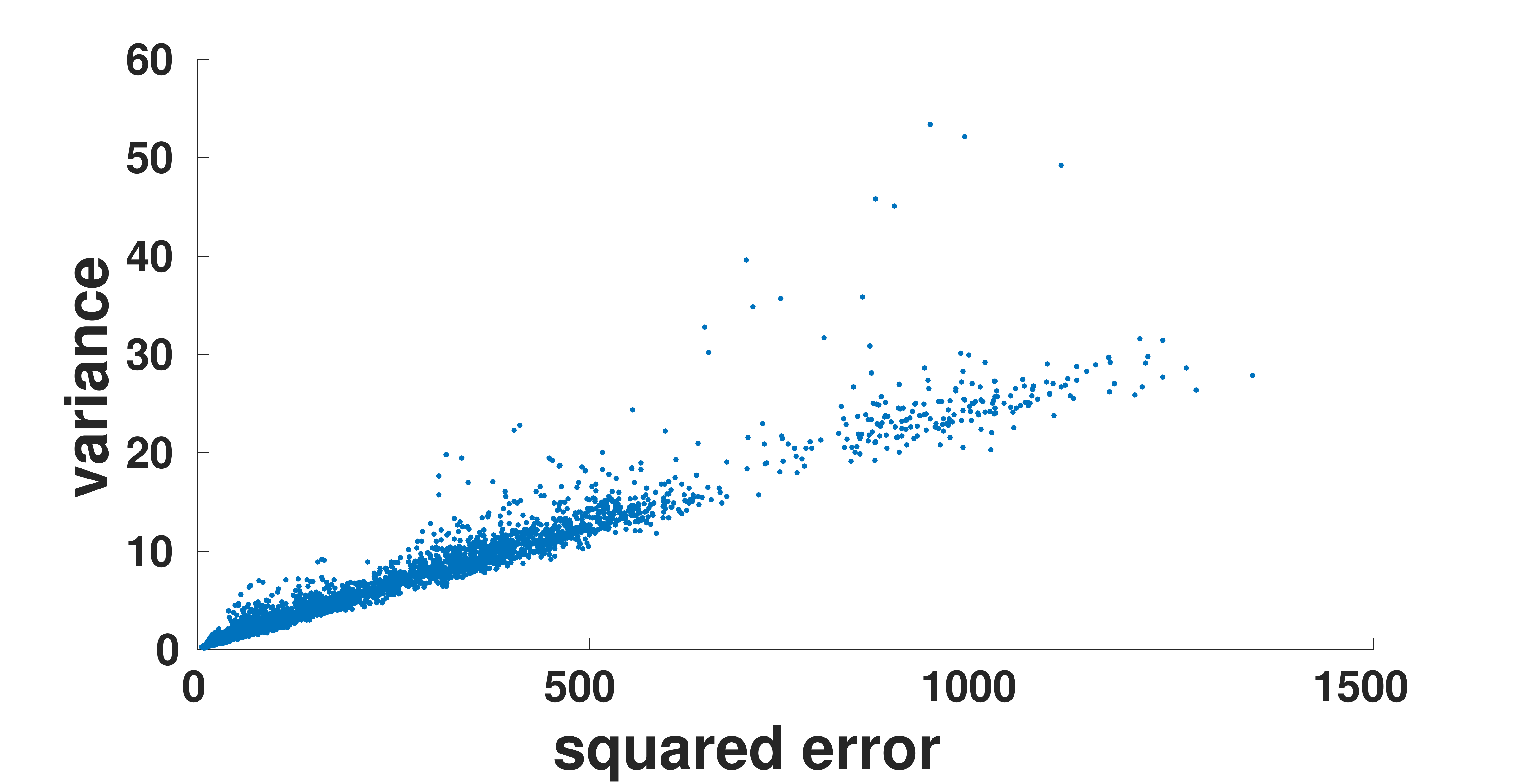}
		&\hspace{-5em}\IncG[width=.2\textwidth,height=.1\textwidth]{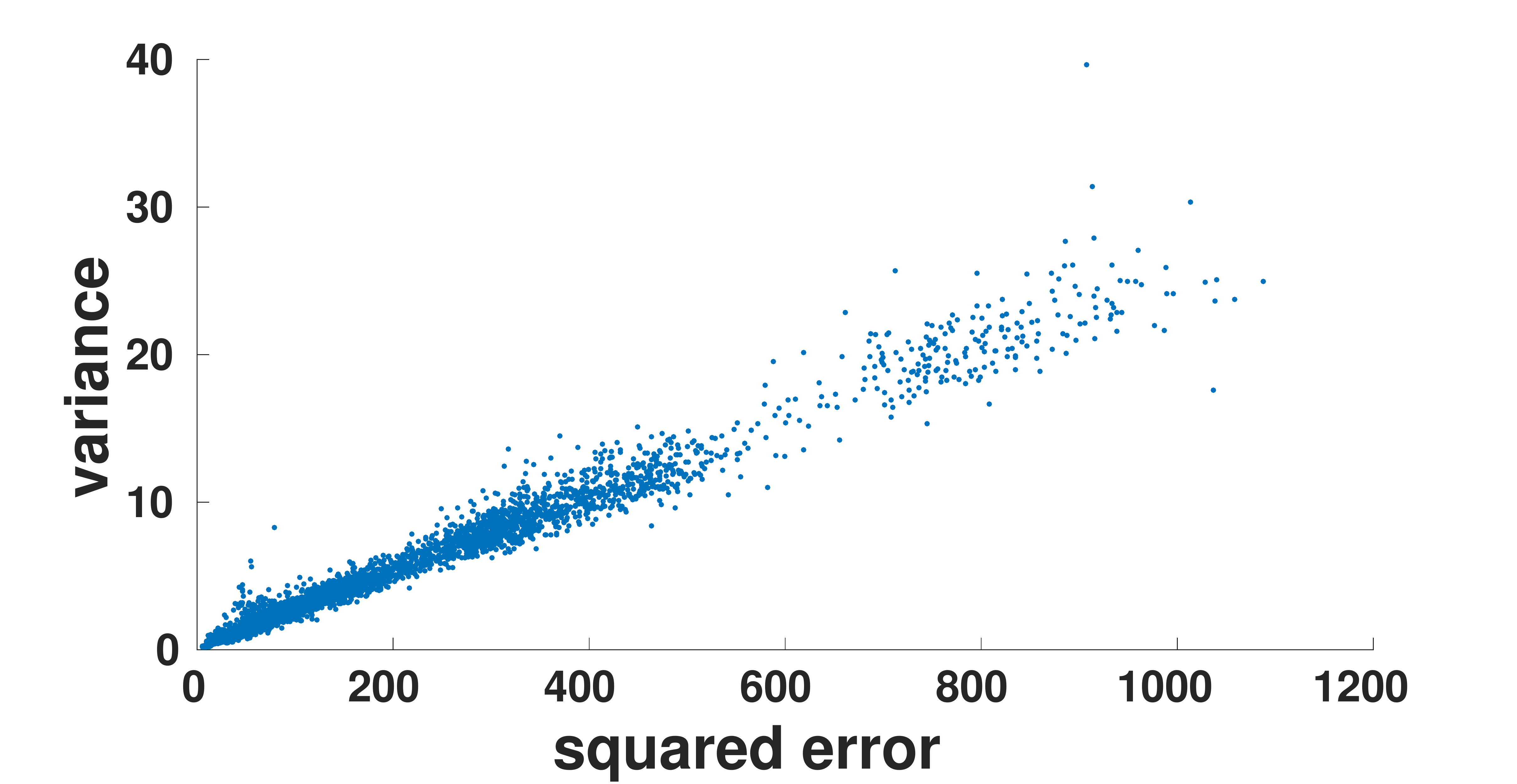}
		&\hspace{-6em}\IncG[width=.2\textwidth,height=.1\textwidth]{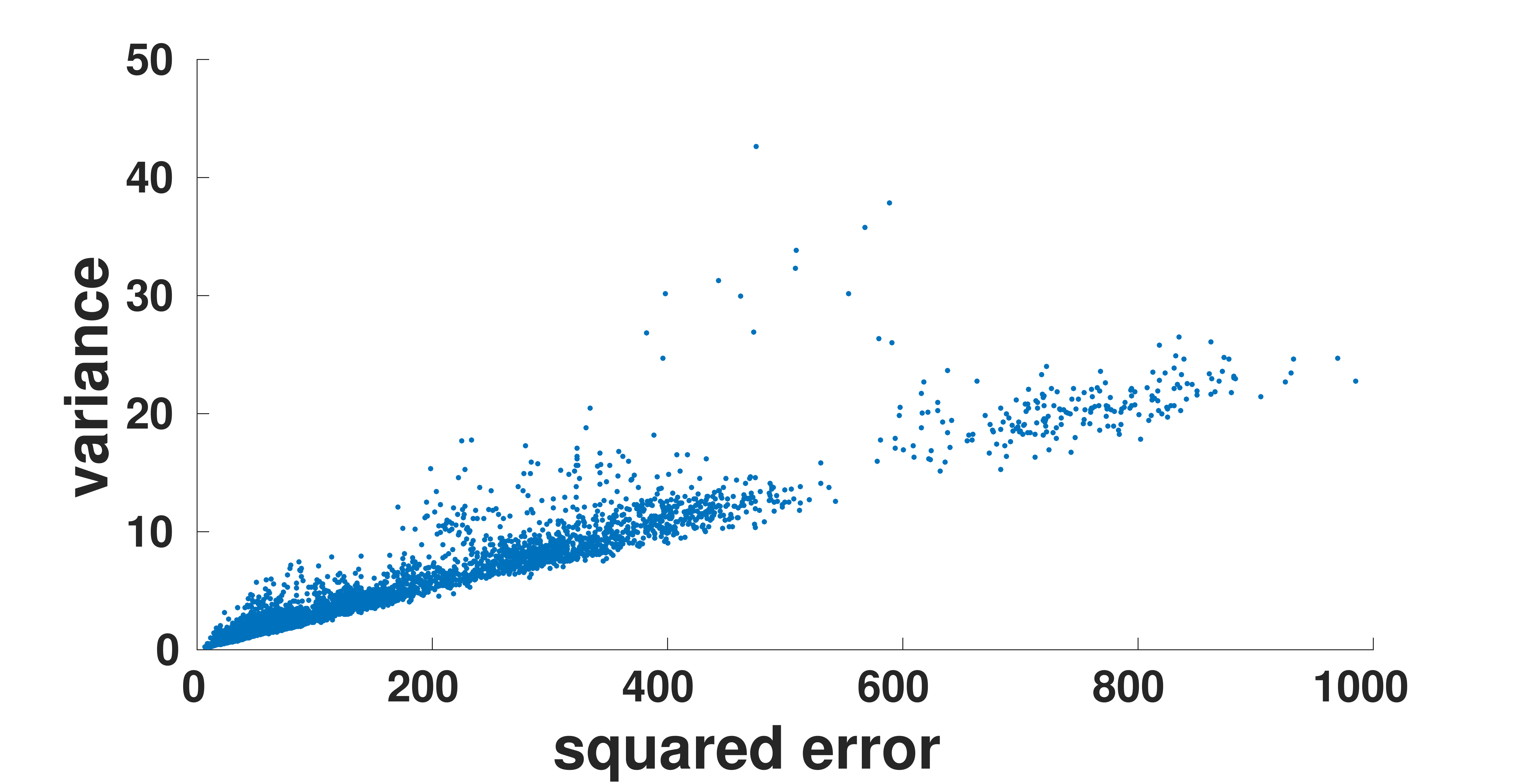}\\
		-5dB 
		&\hspace{-1.5em}\IncG[width=.2\textwidth,height=.1\textwidth]{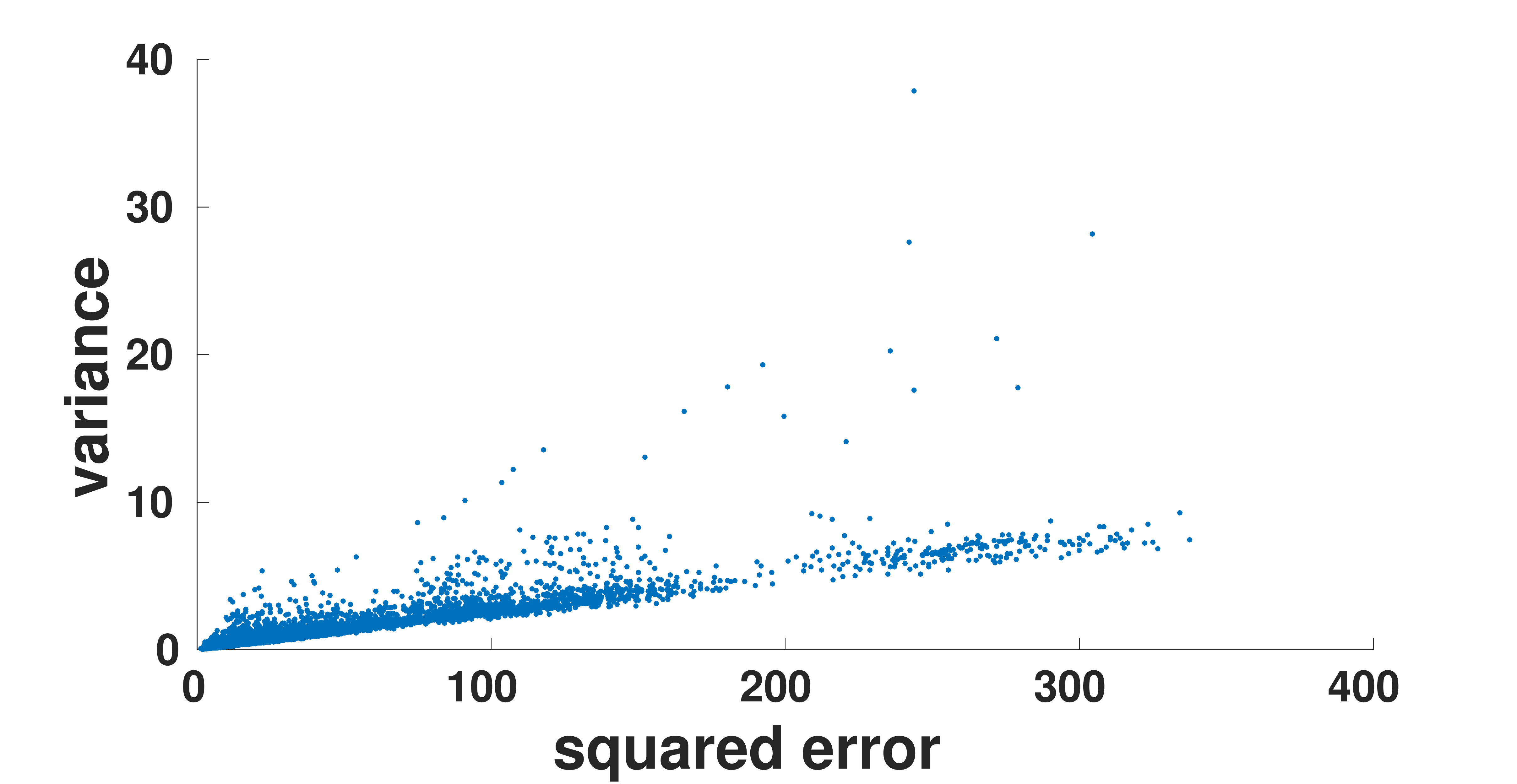}
		&\hspace{-3em}\IncG[width=.2\textwidth,height=.1\textwidth]{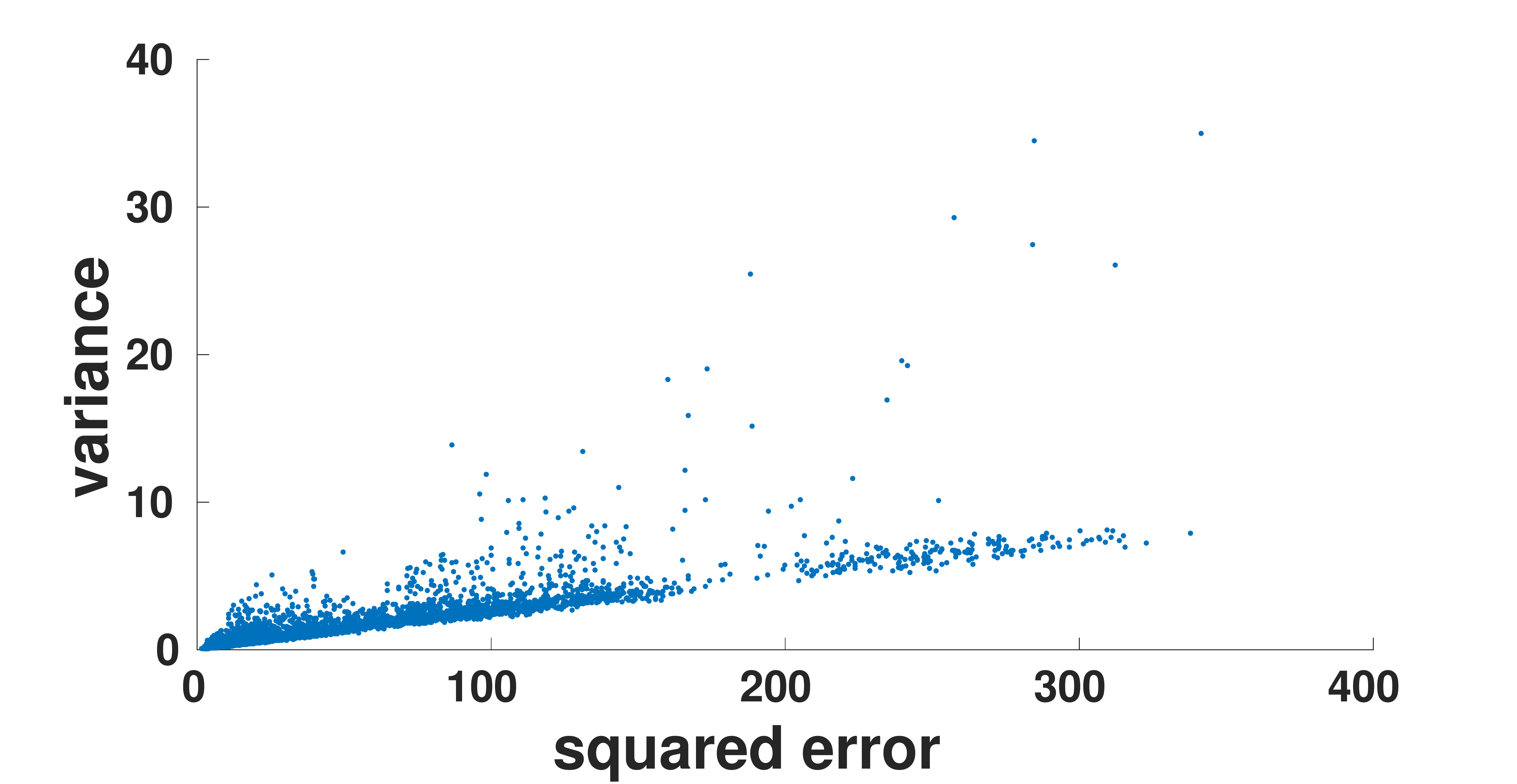}
		&\hspace{-4em}\IncG[width=.2\textwidth,height=.1\textwidth]{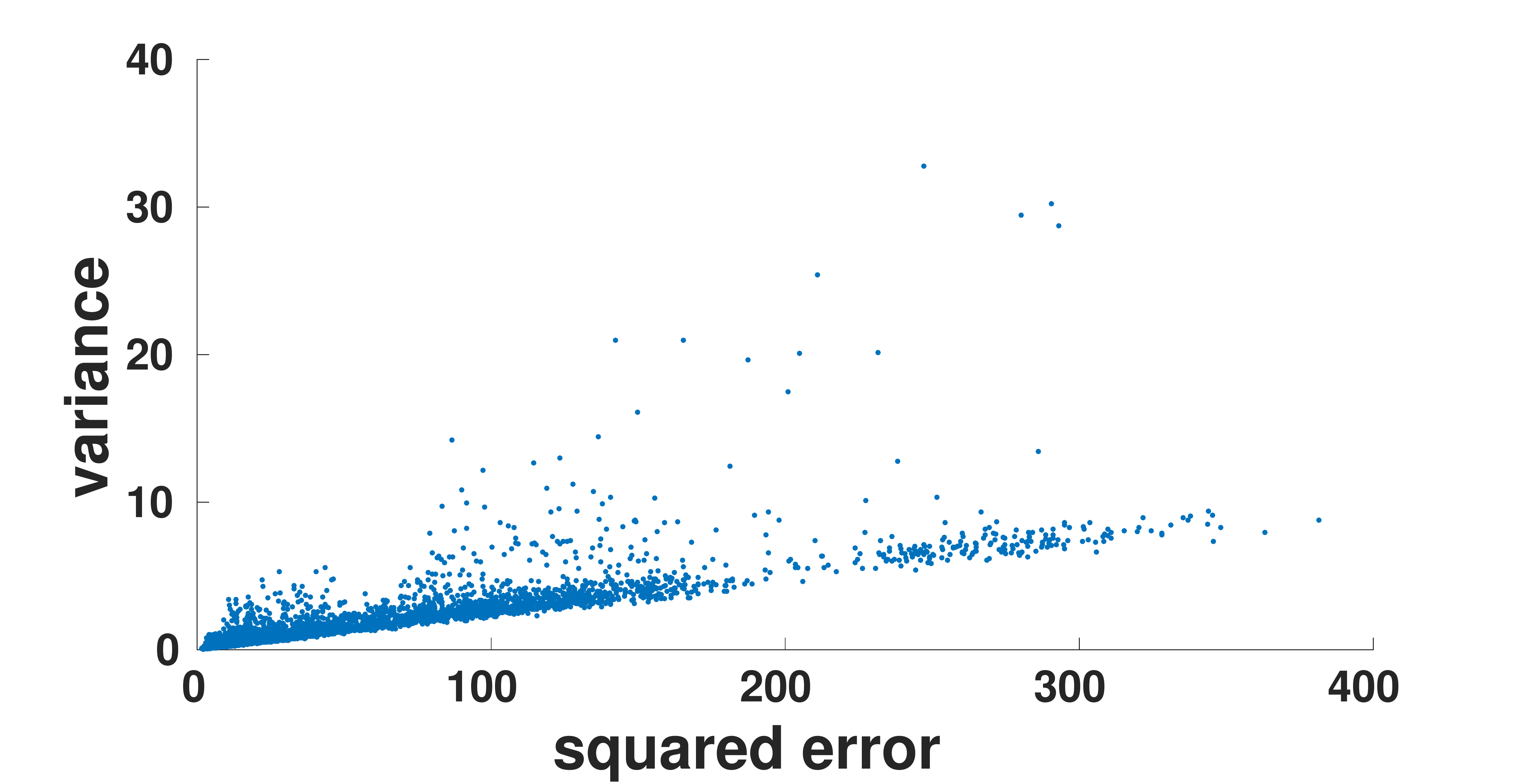}
		&\hspace{-5em}\IncG[width=.2\textwidth,height=.1\textwidth]{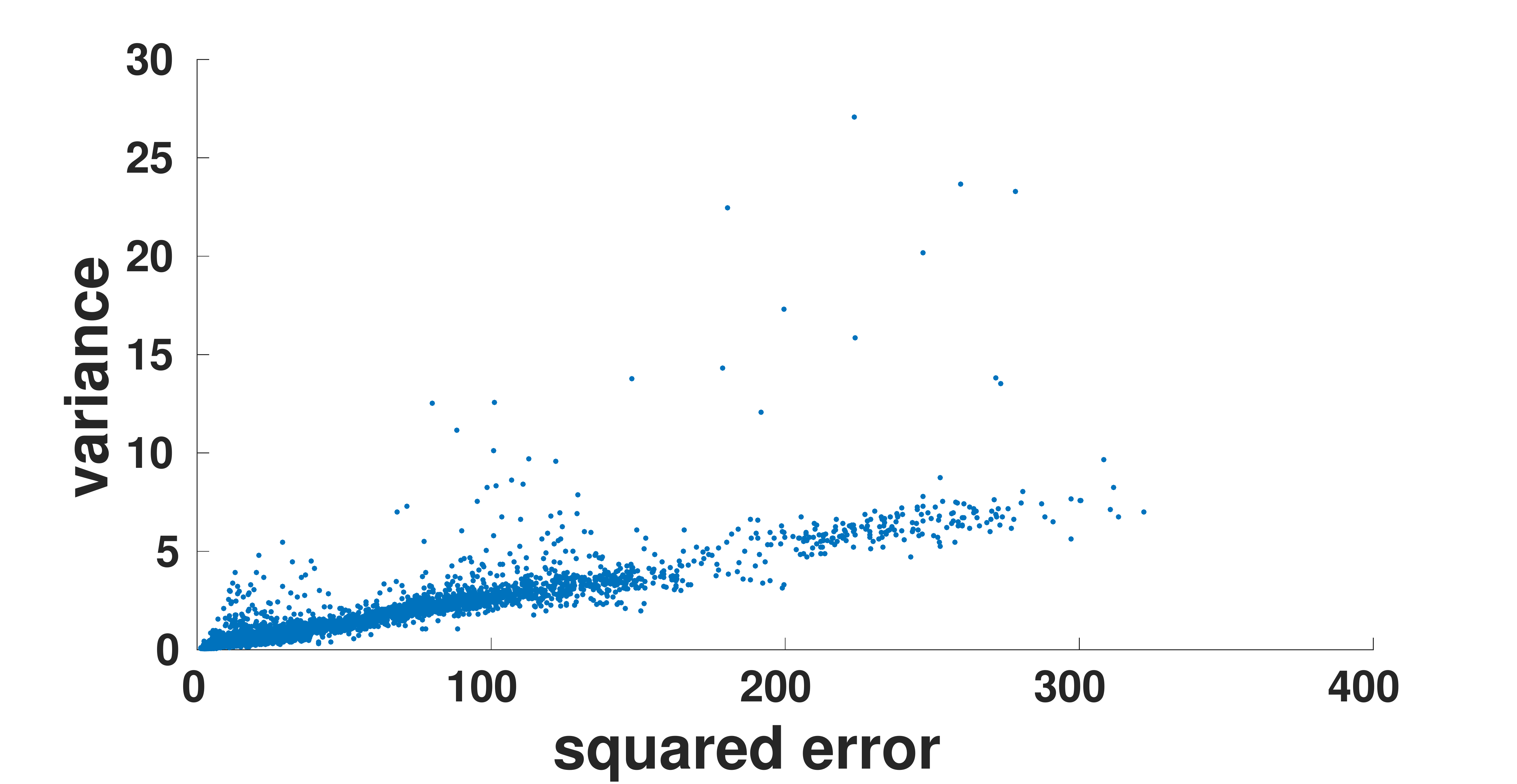}
		&\hspace{-6em}\IncG[width=.2\textwidth,height=.1\textwidth]{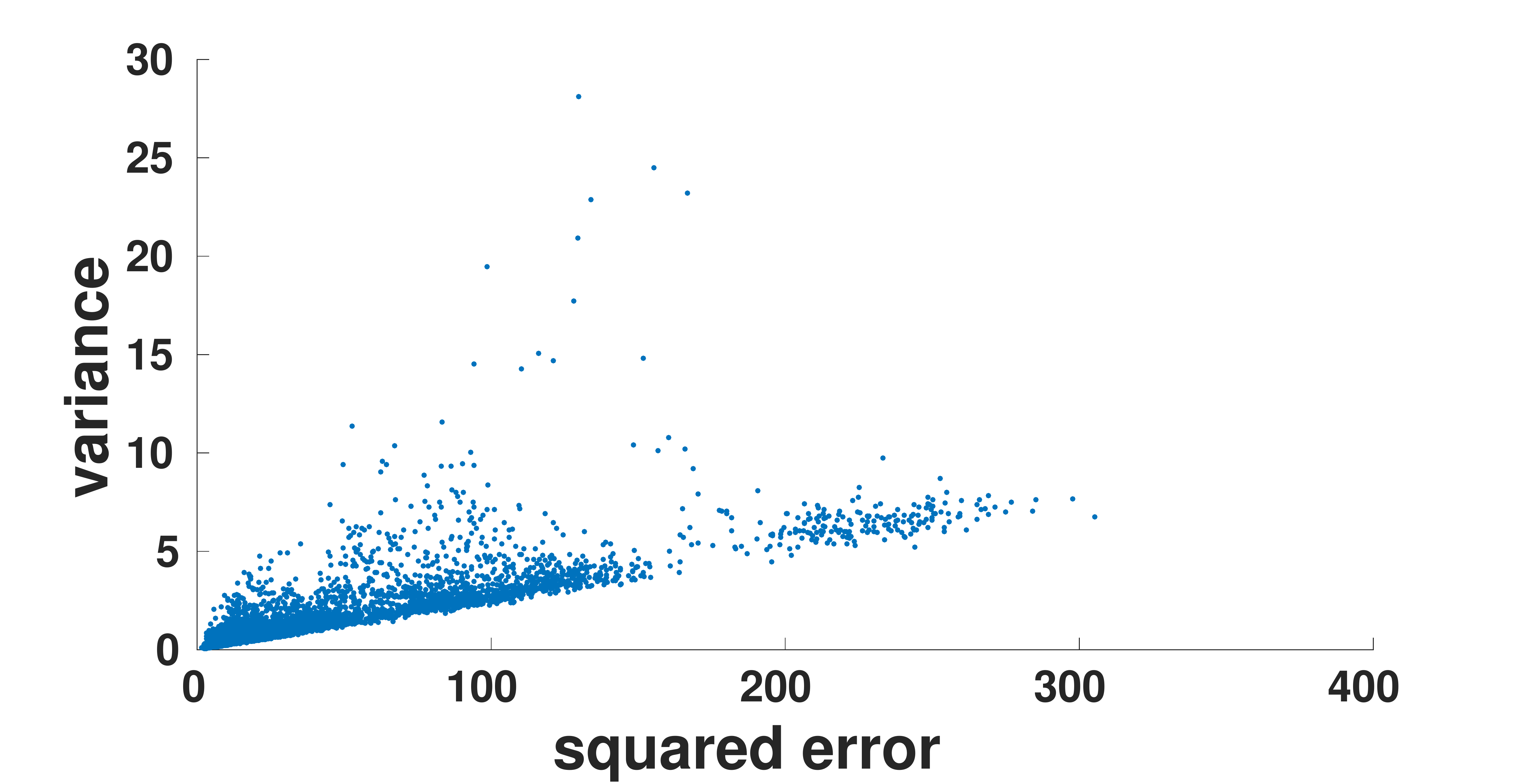}\\
		0dB 
		&\hspace{-1.5em}\IncG[width=.2\textwidth,height=.1\textwidth]{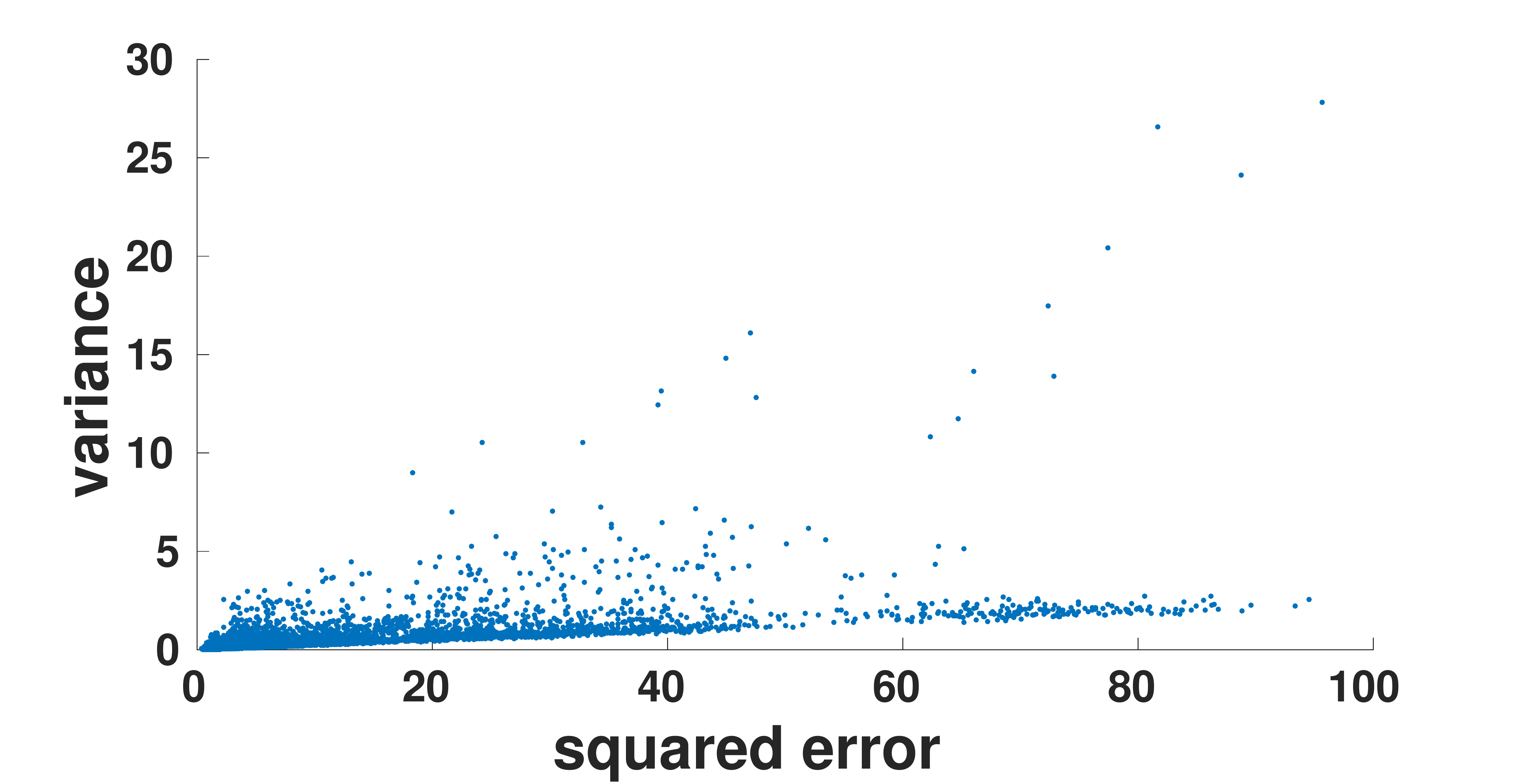}
		&\hspace{-3em}\IncG[width=.2\textwidth,height=.1\textwidth]{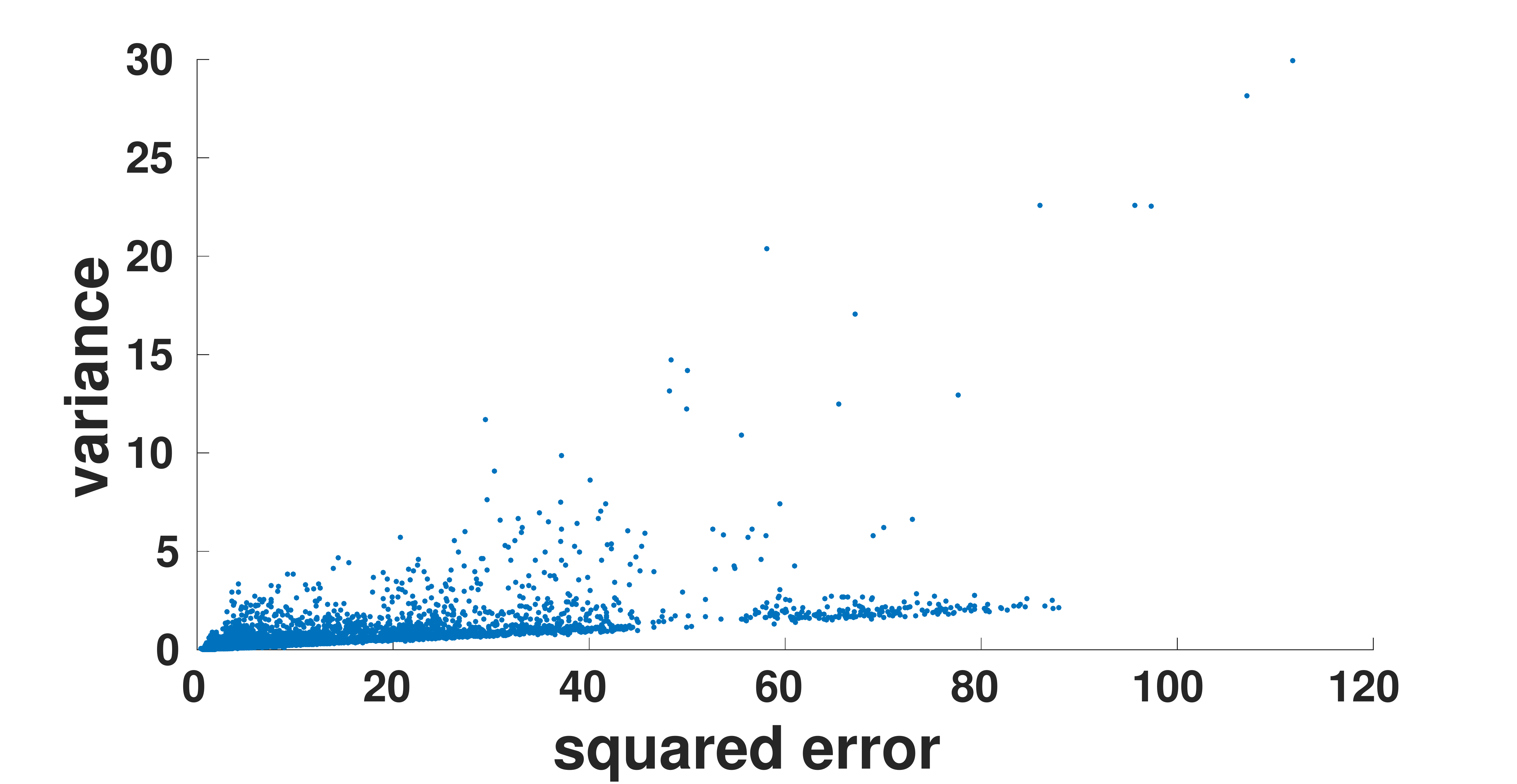}
		&\hspace{-4em}\IncG[width=.2\textwidth,height=.1\textwidth]{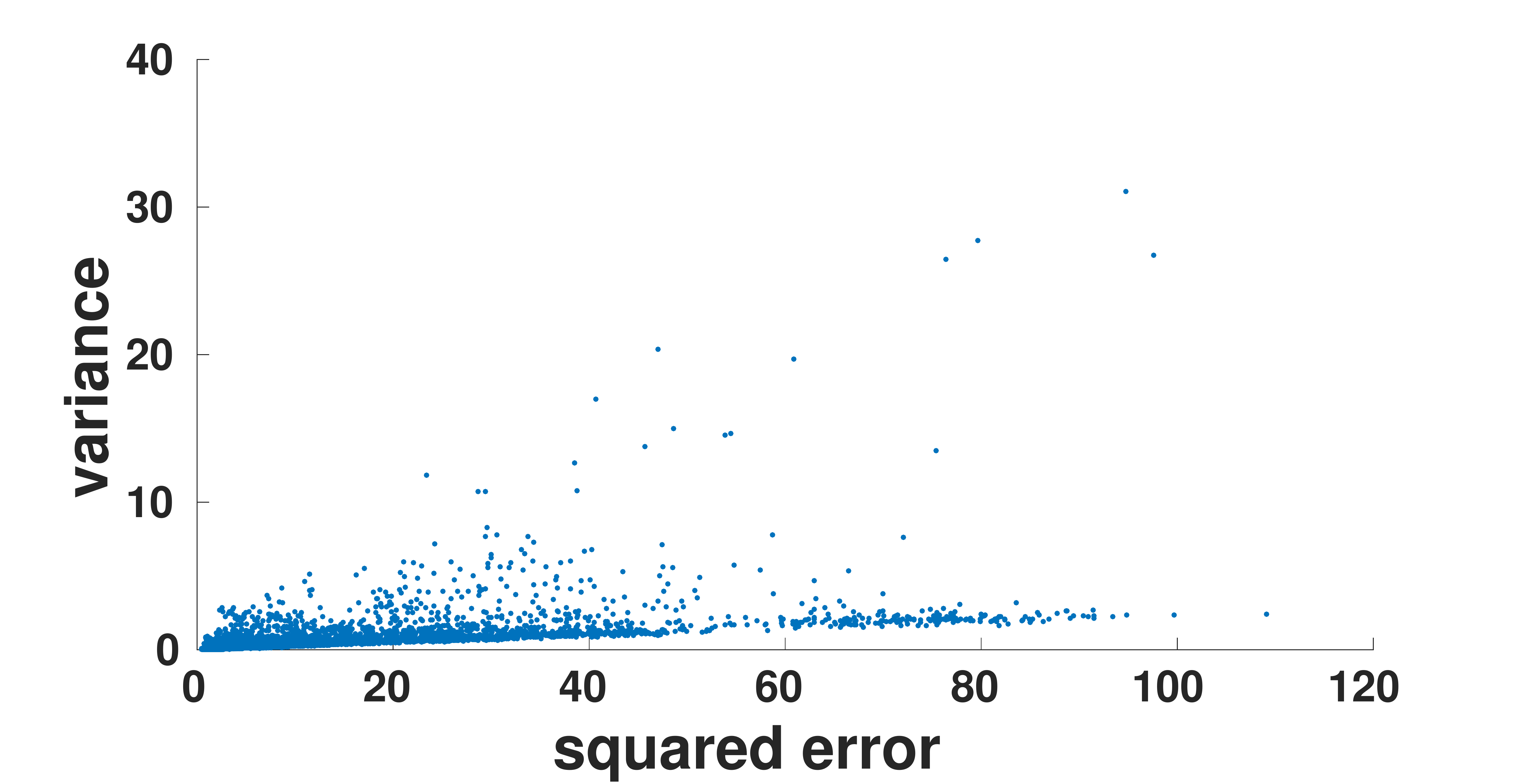}
		&\hspace{-5em}\IncG[width=.2\textwidth,height=.1\textwidth]{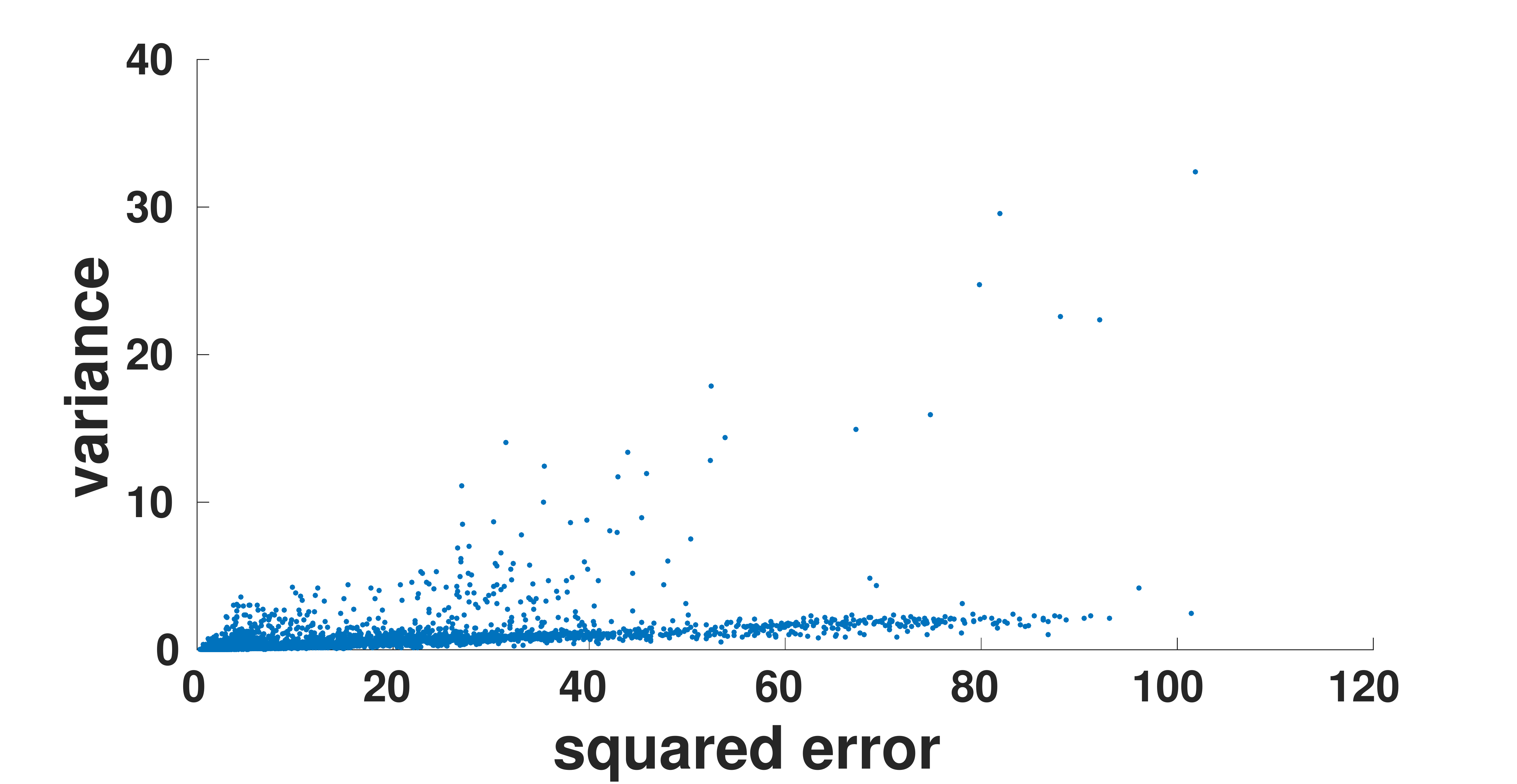}
		&\hspace{-6em}\IncG[width=.2\textwidth,height=.1\textwidth]{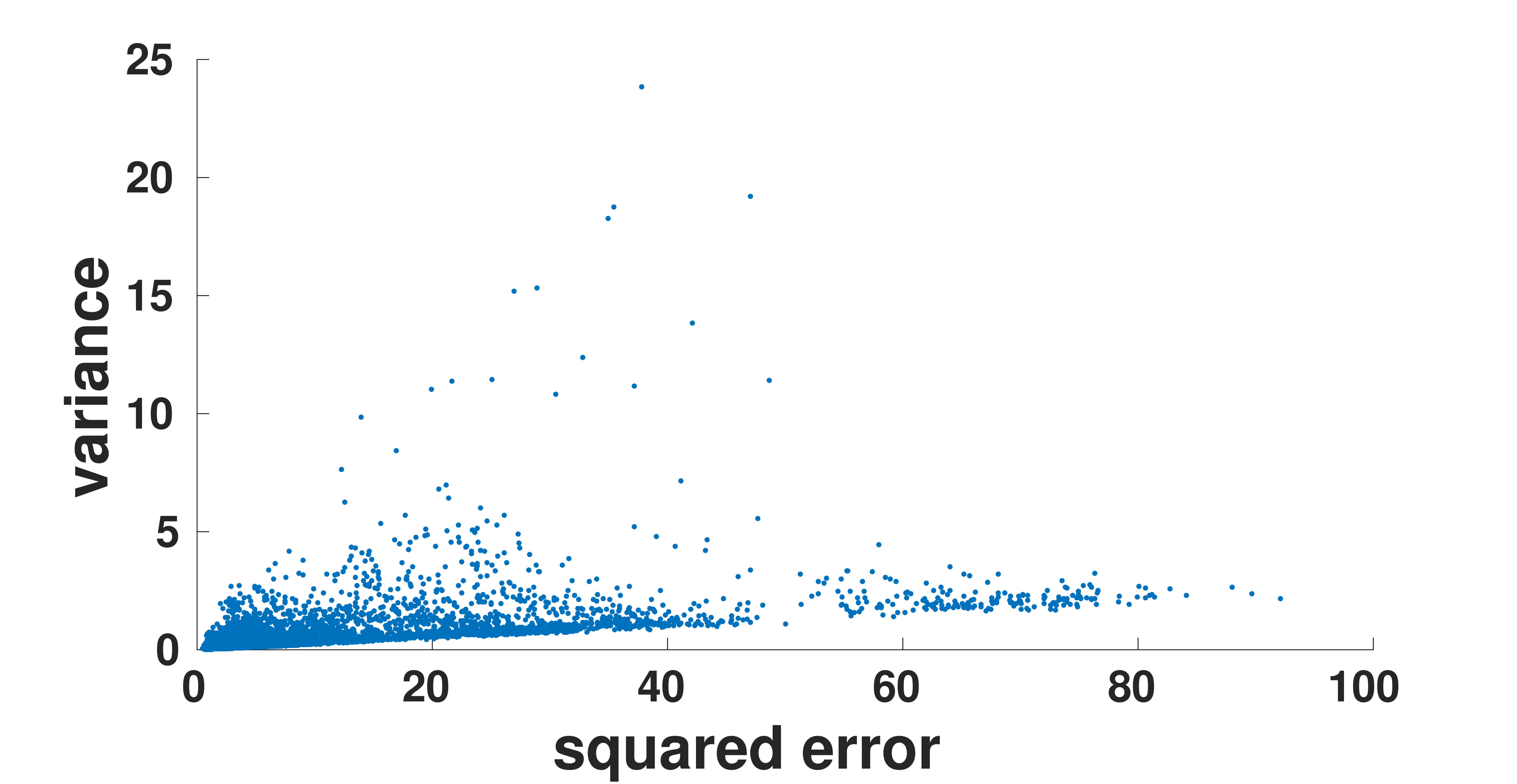}\\
		5dB 
		&\hspace{-1.5em}\IncG[width=.2\textwidth,height=.1\textwidth]{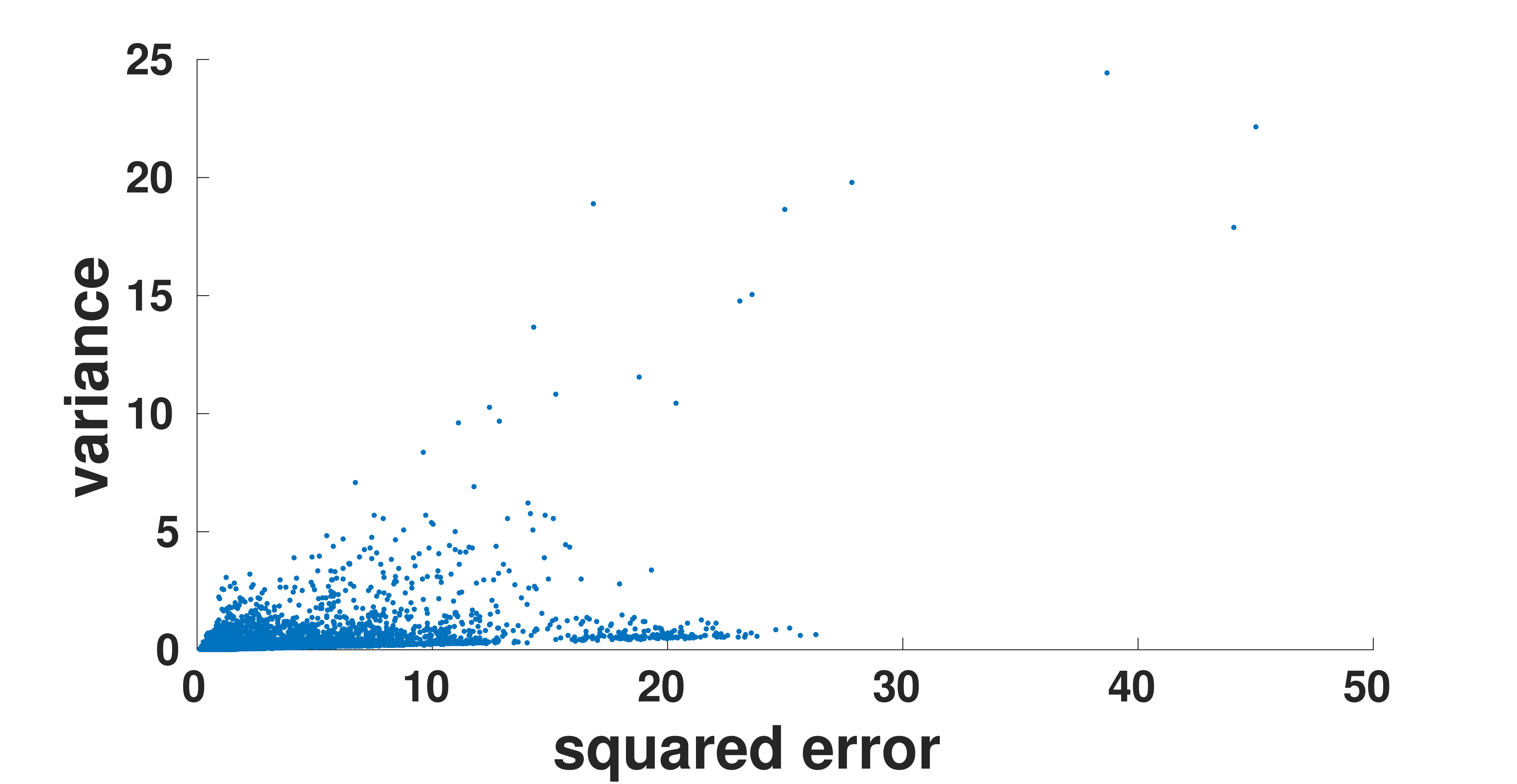}
		&\hspace{-3em}\IncG[width=.2\textwidth,height=.1\textwidth]{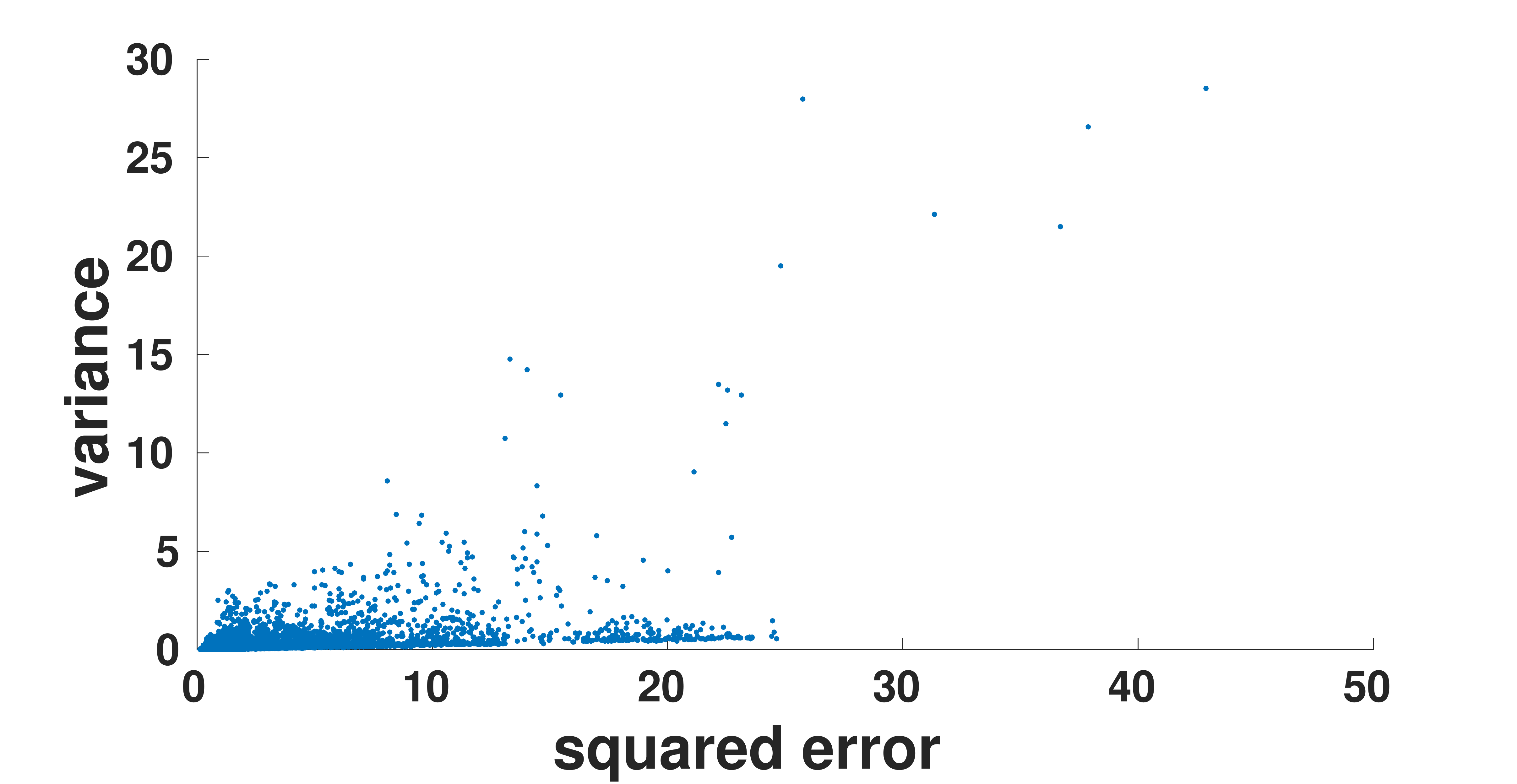}
		&\hspace{-4em}\IncG[width=.2\textwidth,height=.1\textwidth]{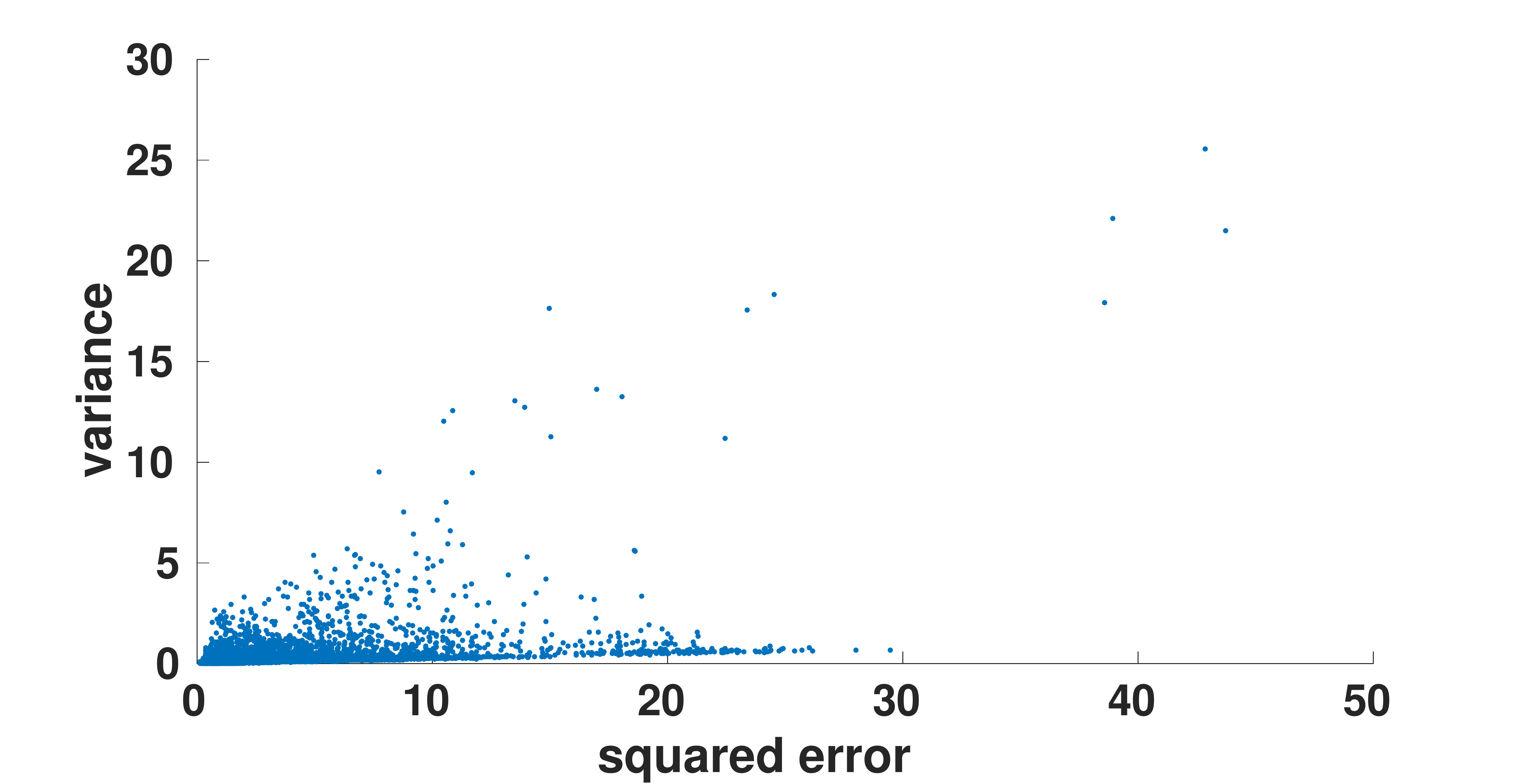}
		&\hspace{-5em}\IncG[width=.2\textwidth,height=.1\textwidth]{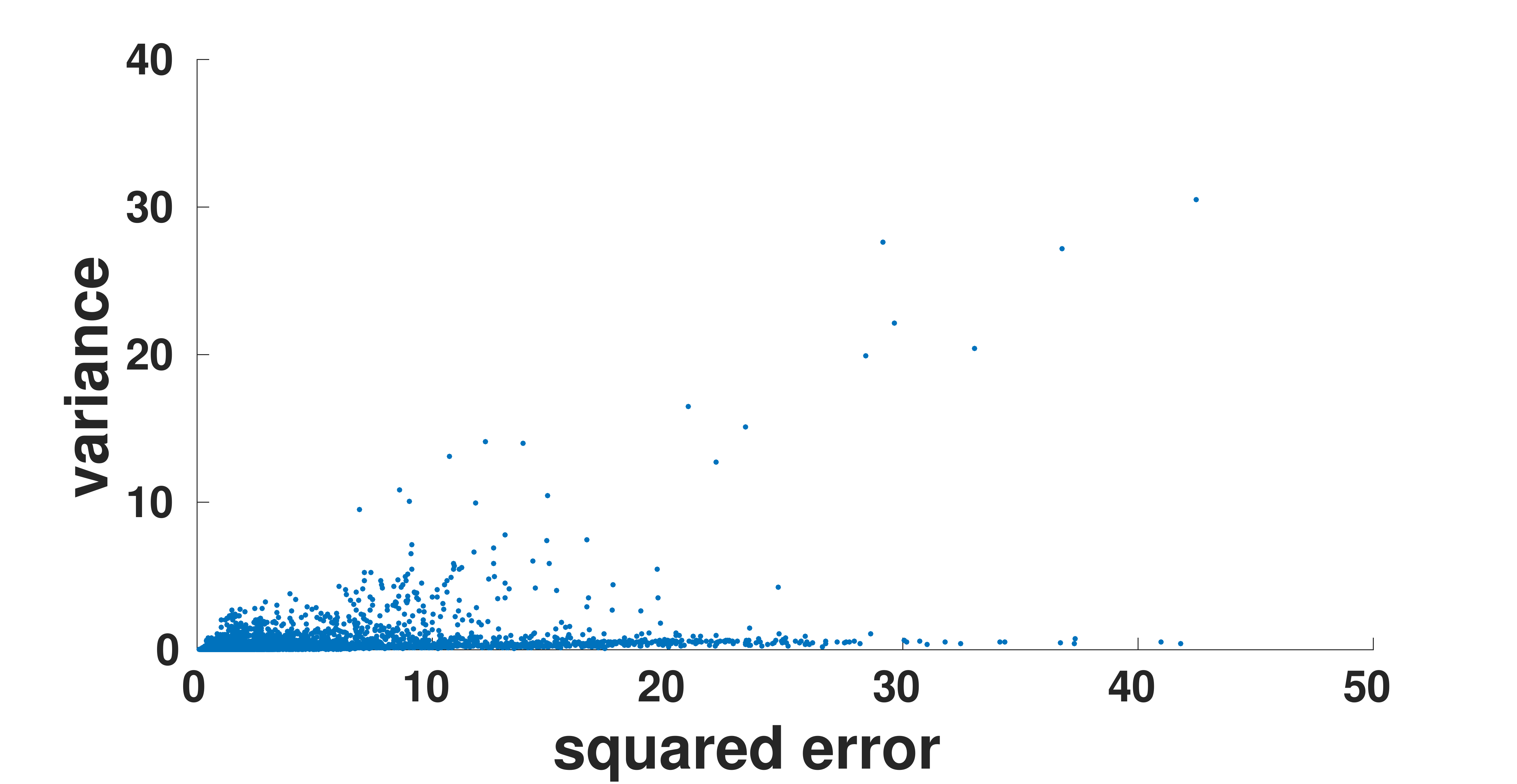}
		&\hspace{-6em}\IncG[width=.2\textwidth,height=.1\textwidth]{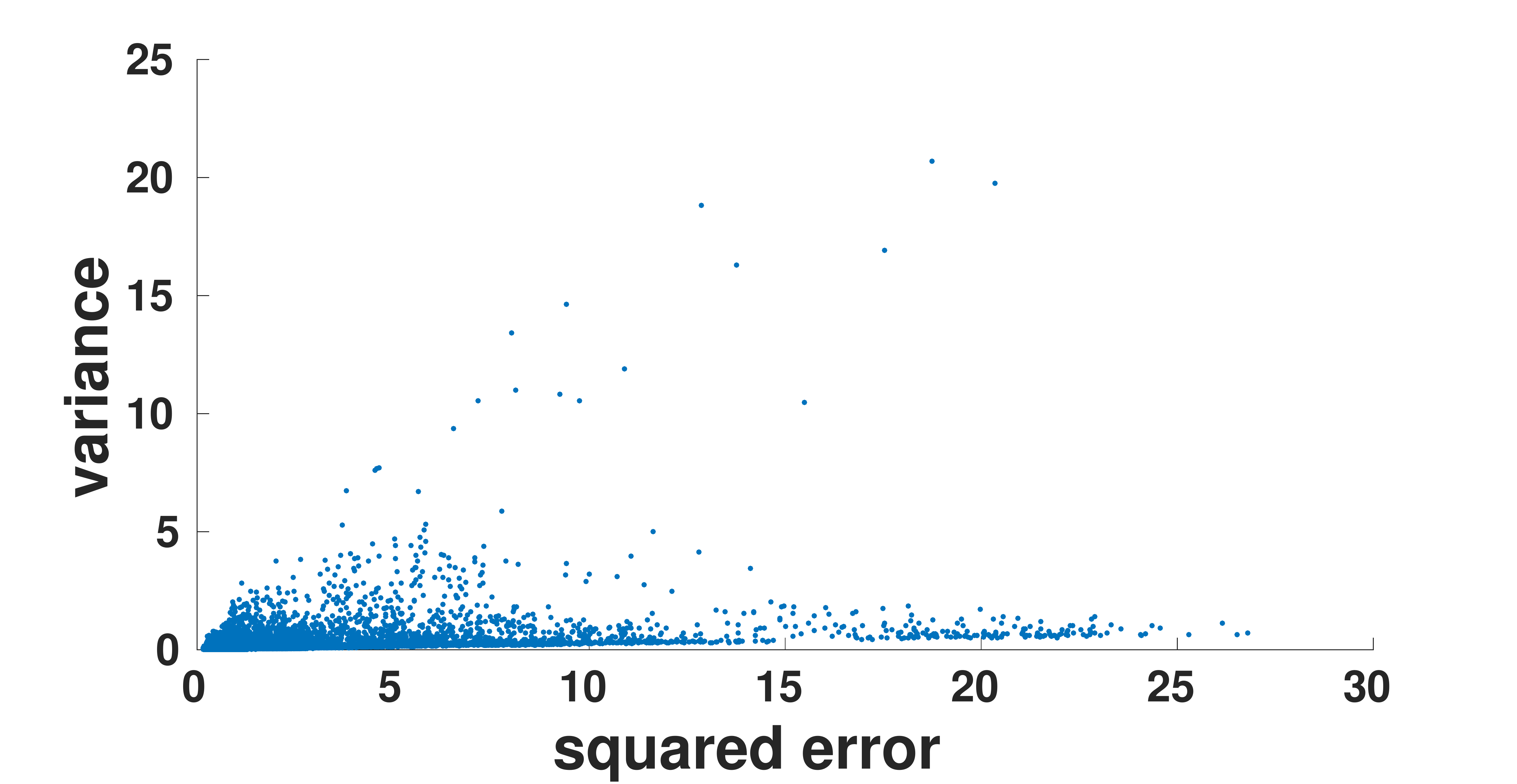}\\
		10dB 
		&\hspace{-1.5em}\IncG[width=.2\textwidth,height=.1\textwidth]{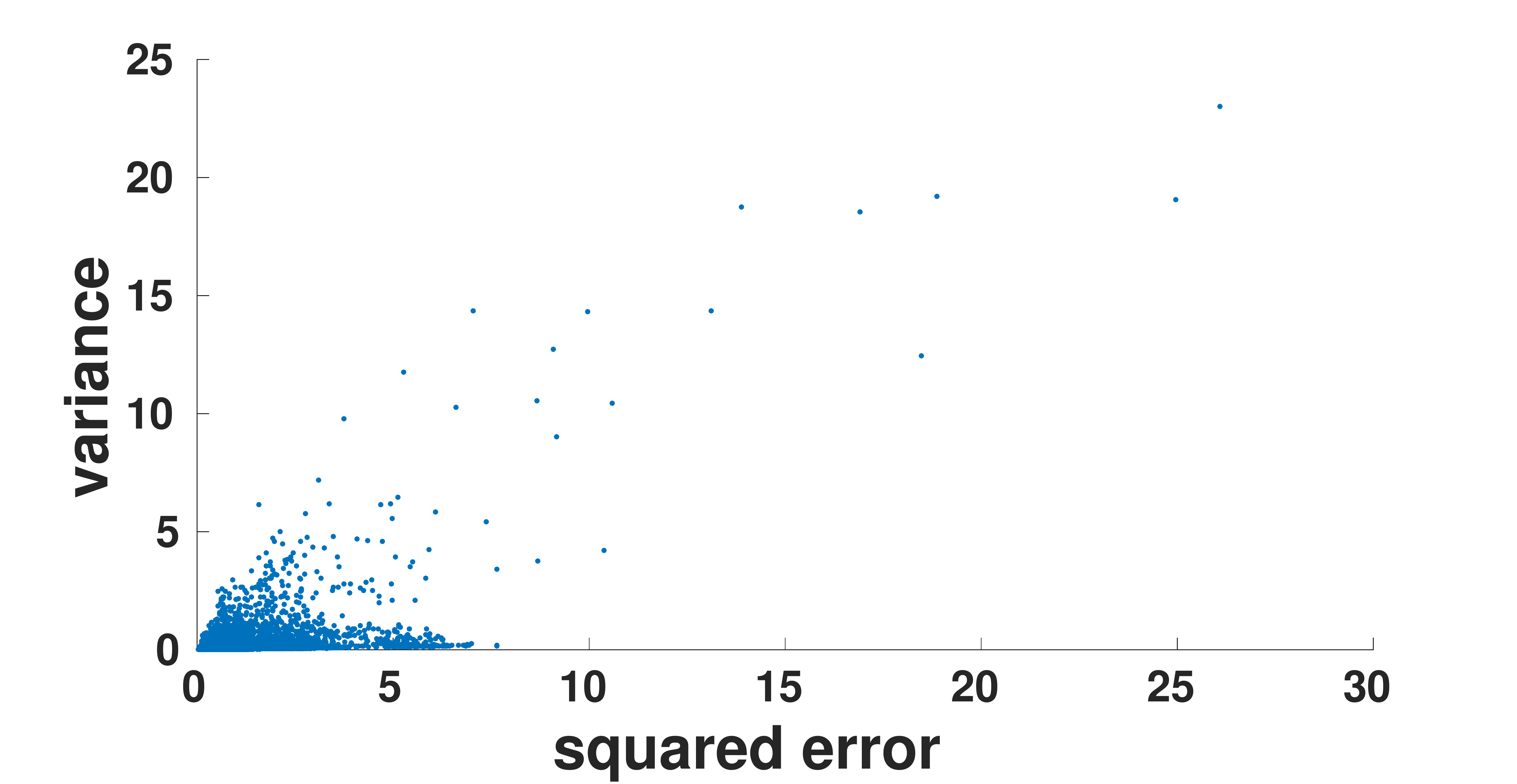}
		&\hspace{-3em}\IncG[width=.2\textwidth,height=.1\textwidth]{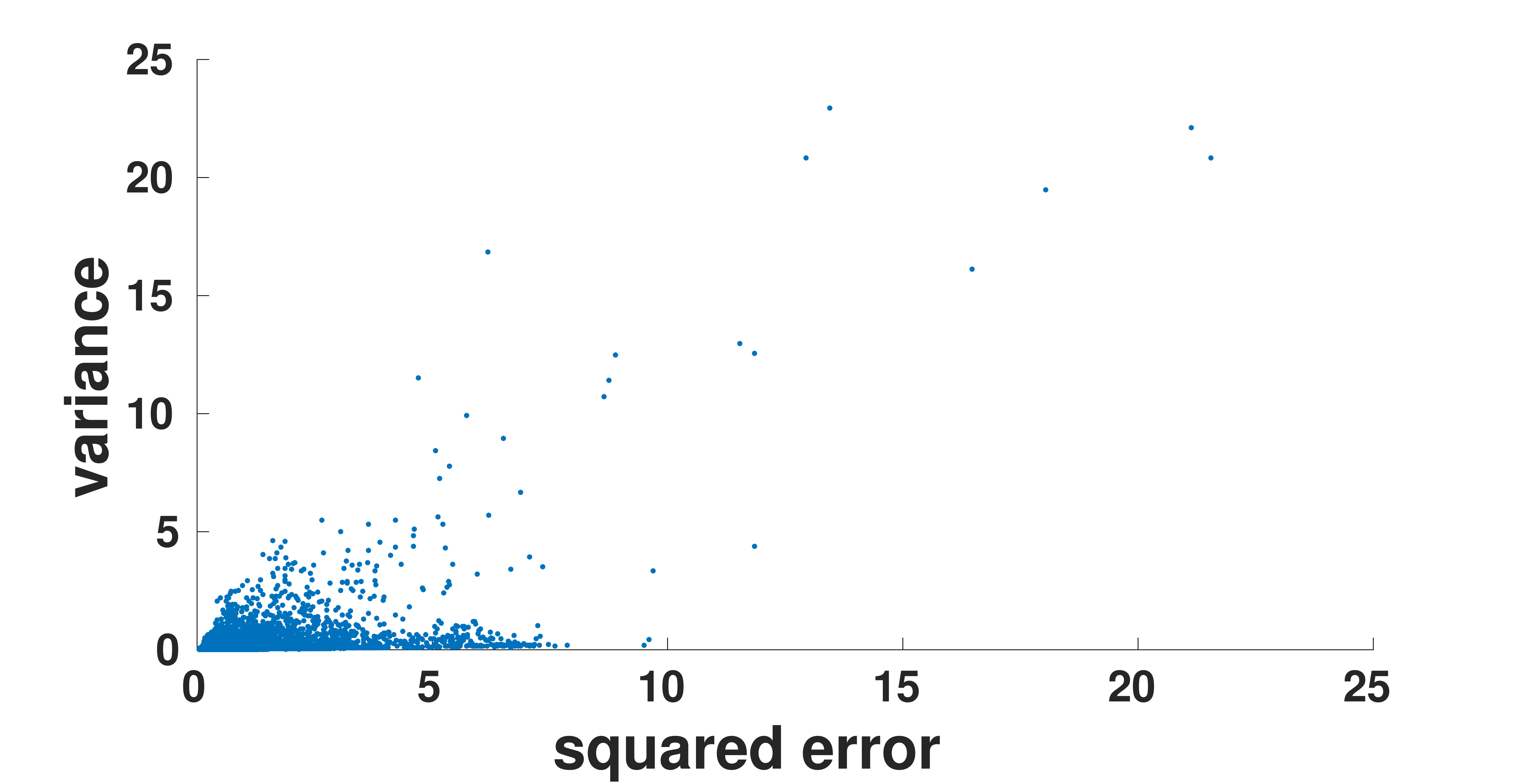}
		&\hspace{-4em}\IncG[width=.2\textwidth,height=.1\textwidth]{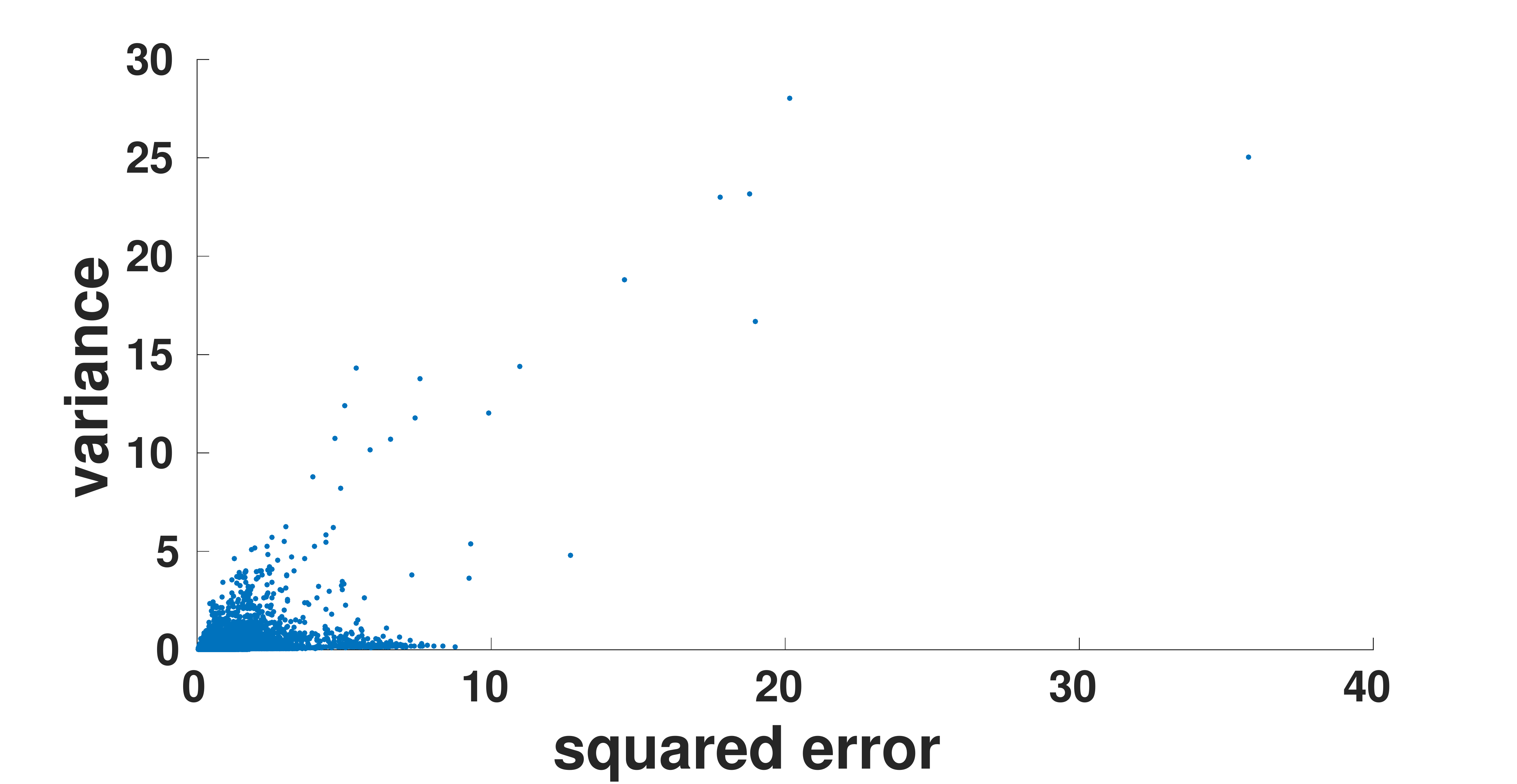}
		&\hspace{-5em}\IncG[width=.2\textwidth,height=.1\textwidth]{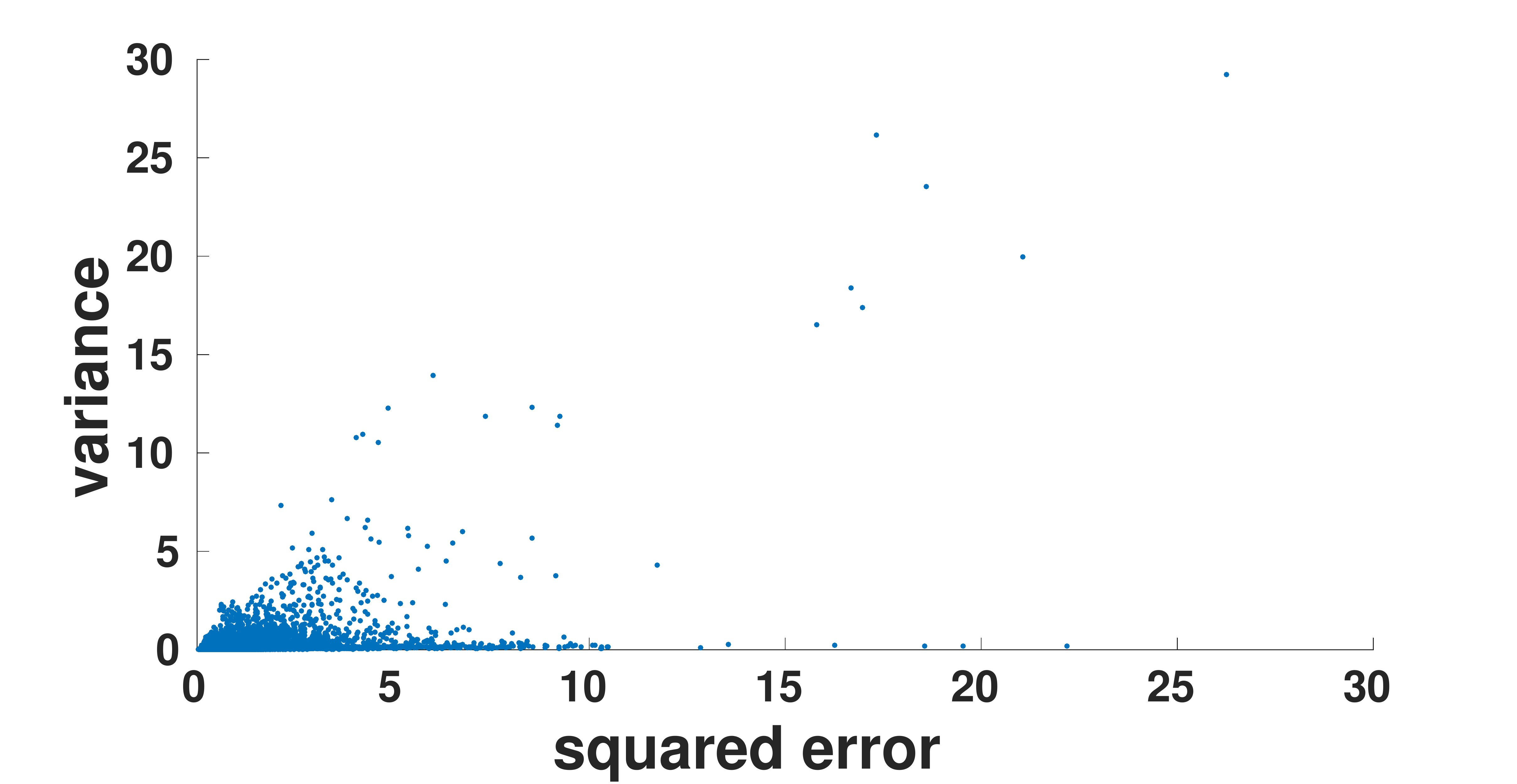}
		&\hspace{-6em}\IncG[width=.2\textwidth,height=.1\textwidth]{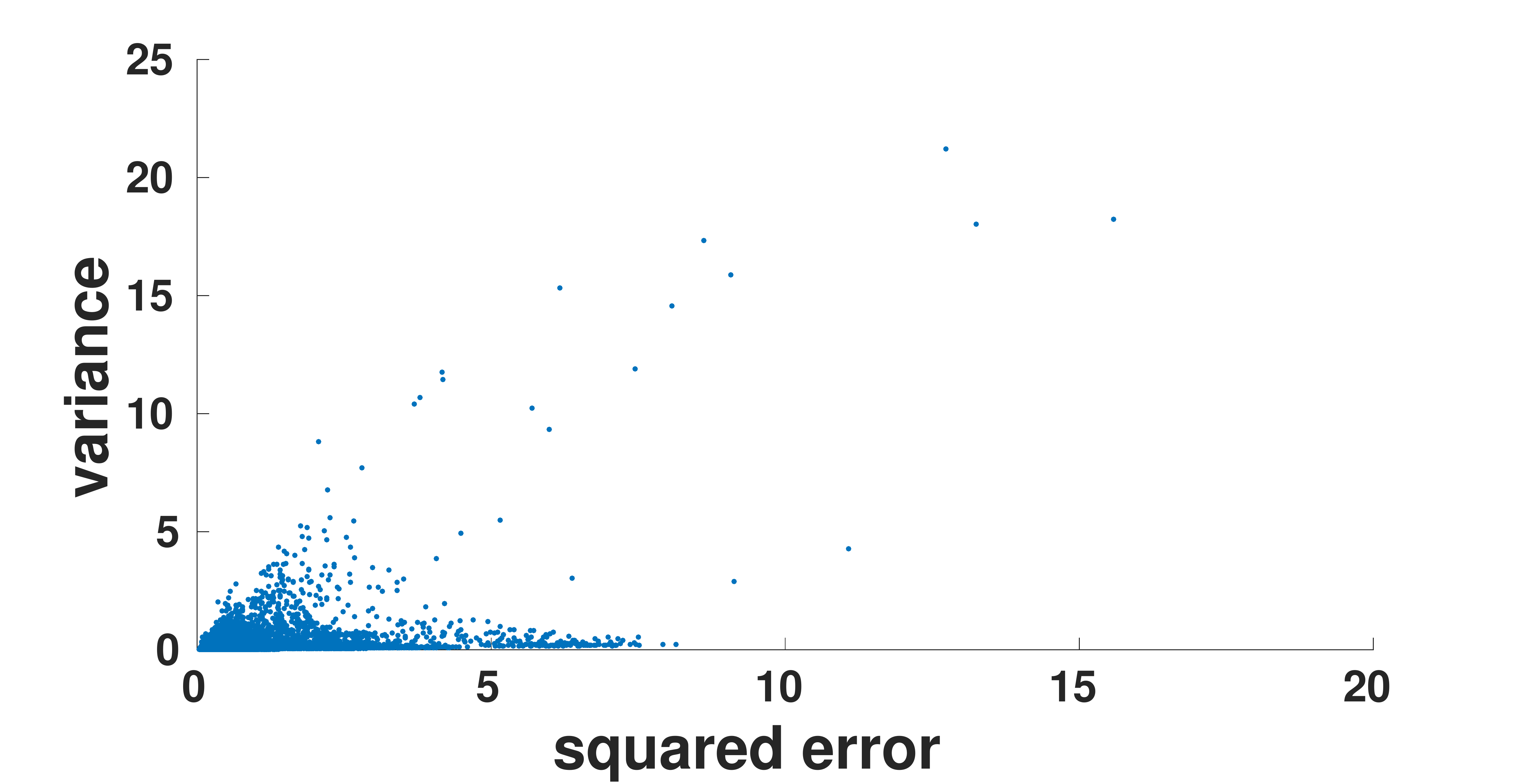}\\
	\end{tabular}

	\vspace{-1em}
	\caption{Correlation plot between the predictive variance and the squared error of the estimated output frames for all the five MC models for the case of speech corrupted with white noise as input}
	\label{fig3}
\end{figure*} 

Table \ref{table2} lists the performance improvements obtained by the multi-model MC dropout DNNs using predictive variance, over the baseline single model in terms of SSE, and SSNR for speech corrupted with unseen noises of white, pink and factory1, averaged over 100 files randomly selected from TIMIT \cite{timit}. Again, the proposed method performs well on low SNRs, especially at -10 dB. However as the SNR improves, the improvement over the baseline drops. This performance drop can be explained by the reduced correlation between the squared error and the model uncertainty that is observed in Fig. \ref{fig3}.  

Figure \ref{fig3} plots the correlation between the predictive variance and the squared error (SE) of the estimated output frames for all the five MC models, for speech with white noise. The uncertainty is computed by taking the trace of the covariance matrix of each frame \cite{kendall2016modelling}. The plots show the weakening of the correlation between  the SE and model uncertainty as the SNR improves. The correlation is strong for -10 and -5 dB and is weak for the values of SNR (0, 5 and 10 dB) on which the model is trained, even if it is with different noises. This matches with our results, since we find that there is not much improvement over the baseline model as the SNR increases. However, the values are still comparable to those of the single model scheme.

From the results in Table \ref{table2} and Fig. \ref{fig3}, the uncertainty based model selection shows promise of being potentially useful, especially in those cases, where the correlation between the model uncertainty and the square error is strong. The interesting pattern on correlation needs further analysis that explores the varying strength of correlation. We would also like to learn the relationship between correlation and squared error better, so that the model can be selected in a risk minimization paradigm. Each model can be trained on a different group of noises and still this algorithm has the potential to be useful.
\vspace{-0.2cm}
\section{Conclusion and future work}
\vspace{-0.1cm}
In this work, we propose two novel techniques that use dropouts as a Bayesian estimator to improve generalizability of DNN based speech enhancement algorithms.  The first method uses the empirical mean of multiple stochastic passes through the DNN-MC dropout model to obtain the enhanced output. Our experiments show that this technique results in superior enhancement performances, especially on unseen noise and SNR conditions. The second method looks at the potential application of the model uncertainty as an estimate of squared error (SE), for frame-wise model selection in speech enhancement, using multiple DNN models. While the experiments on validating this technique give only marginal improvement in some cases, the pattern of correlation between SE and model uncertainty, calls for further study. A particularly interesting line of study would include using complex functions that use the model uncertainty to arrive at the optimal model for each frame. 

\bibliography{asd2_ref.bib}
\bibliographystyle{IEEEtran}

\end{document}